\documentclass[english,prb,showkeys,superscriptaddress,floatfix,twocolumn]{revtex4-2}
\usepackage{ae,aecompl}
\usepackage[T1]{fontenc}
\usepackage[utf8]{inputenc}
\usepackage{babel}
\usepackage{float}
\usepackage{amsmath}
\usepackage{amssymb}
\usepackage{graphicx}
\usepackage{wasysym}
\usepackage{color}
\usepackage[normalem]{ulem}
\usepackage[unicode=true,bookmarks=true,bookmarksnumbered=false,bookmarksopen=false,breaklinks=false,pdfborder={0 0 1},backref=false,colorlinks=true,linkcolor=magenta,citecolor=blue]{hyperref}
\begin{document}
\global\long\def\ket#1{\left|#1\right\rangle }%
\global\long\def\bra#1{\left\langle #1\right|}%
\global\long\def\braket#1#2{\langle#1|#2\rangle}%
\global\long\def\expectation#1#2#3{\langle#1|#2|#3\rangle}%
\global\long\def\average#1{\langle#1\rangle}%

\title{Linear-scale simulations of quench dynamics}
\author{Niaz Ali Khan}
\address{Department of Physics, Zhejiang Normal University, Jinhua 321004, P. R. China}
\affiliation{Zhejiang Institute of Photoelectronics and Zhejiang Institute for Advanced Light Source, Zhejiang Normal University, Jinhua 321004, P. R. China}
\author{Wen Chen}
\address{Beijing Computational Science Research Center, Beijing 100193, China}
\author{Munsif Jan}
\address{Department of Physics, Zhejiang Normal University, Jinhua 321004, P. R. China}
\author{Gao Xianlong}
\email{gaoxl@zjnu.edu.cn}
\address{Department of Physics, Zhejiang Normal University, Jinhua 321004, P. R. China}
\begin{abstract}
The accurate description and robust computational modeling of the nonequilibrium properties of quantum systems remain a challenge in condensed matter physics. In this work, we develop a linear-scale computational simulation technique for the non-equilibrium dynamics of quantum quench systems. In particular, we report a polynomial-expansion of the Loschmidt echo to describe the dynamical quantum phase transitions of noninteracting quantum quench systems. An expansion-based method allows us to efficiently compute the Loschmidt echo for infinitely large systems without diagonalizing the system Hamiltonian. To demonstrate its utility, we highlight quantum quenching dynamics under tight-binding quasicrystals and disordered lattices in one spatial dimension. In addition, the role of the wave vector on the quench dynamics under lattice models is addressed. We observe wave vector-independent dynamical phase transitions in self-dual localization models.
\end{abstract}
\maketitle
\section{Introduction}
\label{sec:intro}
The physical behavior of electronic systems depends on the eigenvalue distribution and the properties of the eigenstates of the Hamiltonian matrix of the system \cite{Wilkinson1988}. The eigendecomposition requires a memory of the order of $\mathcal{O}(N^{2})$ and a computational time scale of $\mathcal{O}(N^{3})$ for an $N-$dimensional dense matrix \cite{Wilkinson1988}. These numerical calculations are computationally expensive, limited to small system sizes \cite{Yin2018,Fadel2021}, and hinder technological applications. The complexity of the quantum properties of most materials requires large-scale computer simulations.

Computational modeling has become an indispensable tool for the calculation of physical quantities, especially on a large scale in condensed matter physics \cite{Weibe2006,Simao2020,Niaz2021,FAN2021,Simao2022,BJELCIC2022108477,Sobczyk2022,Castro2023,Li2023,Zhao2023}. Understanding the quantum properties of strongly correlated and quantum materials has been significantly enriched by the implementation of large-scale computer simulations \cite{Elstner1998,Zen2018}. Examples of simulation methods include quantum Monte Carlo (QMC) methods \cite{Zen2018}, density-matrix renormalization group (DMRG) \cite{Schollwock2005}, density functional theory (DFT) \cite{Dreizler1990,Onida2002}, and kernel polynomial method (KPM) \cite{Weibe2006}. DFT is the leading electronic structure, but it is limited to standard exchange-correlation functionals \cite{Dreizler1990}. KPM is a polynomial-expansion-based method commonly used to calculate static thermodynamic quantities, high-resolution spectral densities, and dynamical correlations at zero temperature \cite{Weibe2006,Simao2020}.

The simplest paradigm of nonequilibrium dynamics is a quantum quench process in which an abrupt change in the system parameter independently controls its time evolution \cite{Heyl2015,Heyl2018,Sadrzadeh2021,Jafari2022,Nicola2022}. Recent progress suggests that the quantum quenching process can be used as a theoretical tool to study many aspects of nonequilibrium physics, such as universal aspects of quantum critical dynamics \cite{Hamazaki2021,Damme2022}. The study of dynamical quantum phase transitions (DQPTs) is particularly important as it combines aspects of symmetry, topology, and nonequilibrium physics \cite{Peotta2021}. The quantum quench problem comes in a variety of shapes, including the quantum Ising model \cite{Zurek2005}, the quantum quench of an atomic Mott insulator \cite{Chen2011}, the Aubry--André model with a quenched incommensurate potential \cite{Yang2017,Xu2021}, the Lipkin-Meshkov-Glick model with a quenched transverse field \cite{Xu2020}, the Anderson model with quenched disorder \cite{Yin2018}, the Aubry--André model with a $p-$wave superconducting pairing \cite{Tong2021} and the correlated Anderson model with quenched disorder correlation strength \cite{Niaz2023cDQPT}. Remarkably, quench-induced DQPTs provide a new framework for exploring the dynamical behavior of quantum systems evolving over time \cite{Peotta2021}. The concepts of non-equilibrium quantum criticalities have been mapped onto DQPTs, characterized by singularities in the Loschmidt echo at certain critical times of quenched systems \cite{Yang2017,Xu2021}. A Loschmidt echo is a measure of the overlap between the initial ground state and its evolution in time and has been widely studied both theoretically \cite{Naldesi2016,Wong2022,Xu2020,Yang2017,Tong2021,Xu2021,Niaz2023cDQPT,Niaz2023dDQPT,Zou2023,Vanhala2023} and experimentally \cite{Jurcevic2017,Flaschner2018}.

An exact diagonalization method (EDM) is commonly used for the numerical treatment of Loschmidt echos, which is computationally expensive and limited to small system sizes. To overcome this difficulty, we implement a polynomial-expansion-based technique for large-scale tight-binding simulations of Loschmidt echoes in electronic systems. The polynomial-expansion algorithm reduces the numerical complexity to at most $\mathcal{O}(N)$ by truncating polynomial-expansions, which in turn control the numerical accuracy of the approximation. Importantly, the reduction in computational time allows us to study systems with infinitely large sizes. The polynomial-expansion algorithm has the distinct advantage that, in principle, it gives the correct dynamical phase transition of quenched systems. We demonstrate the implementation of the polynomial-expansion of the Loschmidt echo for the Aubry--André model with a quenched incommensurate potential, the uncorrelated Anderson model with a quenched disorder potential, and the correlated Anderson model with quenched disorder correlations. Our results shed new light on the non-equilibrium critical properties of electronic systems and show the existence of DQPTs under certain conditions. The role of the wave vector in the quench dynamics of the models is also discussed.

The structure of our paper is as follows: Sec. \ref{sec:ModelQuenchDynamics} discusses tight-binding models of one-dimensional (1D) noninteracting spinless fermions in a diagonal potential. We also briefly review the established nonequilibrium transport formalism based on the Loschmidt echo. Sec. \ref{sec:KernelPolynomialMethod} focuses on the numerical simulations of the Loschmidt echo based on the polynomial-expansion technique. Sec. \ref{sec:ApplicationsKPM} deals with the implementation of polynomial-expansions for the Loschmidt echo of quenched tight-binding models with diagonal potential and discusses the dynamical properties of the systems. Sec. \ref{sec:Benchmark} shows the benchmark of the Loschmidt echo simulations for models. The last section summarizes our conclusions.
%
\section{Model and quench dynamics\label{sec:ModelQuenchDynamics}}
This section is devoted to a comprehensive study of 1D electronic systems with different local potentials and discusses the transport properties of the systems in an equilibrium setting. We emphasize the localized or extended nature of the controlling parameters of quantum systems. We also briefly discuss the quench dynamics of the system as probed by the Loschmidt echo.

\subsection{Theoretical models}
The model we focus on consists of noninteracting spinless electrons in a 1D quantum system with nearest-neighbor hoppings. The Hamiltonian of the system has the general form \cite{Moura1998,Niaz2019},
\begin{equation}
\hat{\mathcal{H}}=-t\sum_{n=1}^{N}(c_{n}^{\dagger}c_{n+1}+c_{n+1}^{\dagger}c_{n})+\sum_{n=1}^{N}\varepsilon_{n}c_{n}^{\dagger}c_{n},\label{eq:1DHamiltonian}
\end{equation}
where $c_{n}^{\dagger}$ and $c_{n}$ are the free fermionic creation and annihilation operators at site $n$ respectively. The parameter $\varepsilon_{n}$ denotes the random energy of an electron at the $n$-th site of the lattice of size $N$ and $t$ is the hopping integral (transfer energy) between the nearest neighboring sites. The hopping integral is set to unity, and all the other energy scales are measured in units of $t$, and the periodic boundary condition (PBC) is used. The fermionic system is defined on a one--dimensional lattice consisting of $n = 1, . . . , N$, identical cells, spanned by basis of $N$ single--particle states acting on a fermionic Hilbert space (Fock space) $\mathbf{H}$. In matrix representation, the fermionic system under consideration has a $N \times N$ single particle tight-binding Hamiltonian matrix and acts on a Hilbert space, whose eigenstates can be expanded in terms of the site basis.

For the Aubry--André model \cite{AAModel1980}, the lattice site energy is the quasiperiodic potential given by
\begin{align}
\varepsilon_{n} & =\lambda\cos\left(2\pi\beta n\right),\label{eq:IncommensuratePotentiall}
\end{align}
where $\lambda$ is the strength of the incommensurate diagonal energy with $\beta = (\sqrt{5}-1)/2$, being an irrational number. The system displays metal-insulator transitions at $\lambda = 2t$. It is well established that all eigenstates are localized for $\lambda \geqslant 2t$ and extended for $\lambda \leqslant 2t$.

For the standard Anderson model \cite{Anderson1958}, the lattice site energies $\varepsilon_{n}$ are independent random variables distributed uniformly in the interval $[-W/2, W/2]$. Here, $W$ denotes the width of the distribution, which controls the amplitude of the disorder. It is well known that all eigenstates in non-interacting low-dimensional systems are localized by an infinitesimal amount of disorder in the thermodynamic limit, whereas a three-dimensional system exhibits a metal-insulator transition at critical disorder strength with a mobility edge demarcating extended and localized states \cite{Balian1984,Niaz2021}
\begin{figure}
\begin{centering}
\includegraphics[scale=0.36]{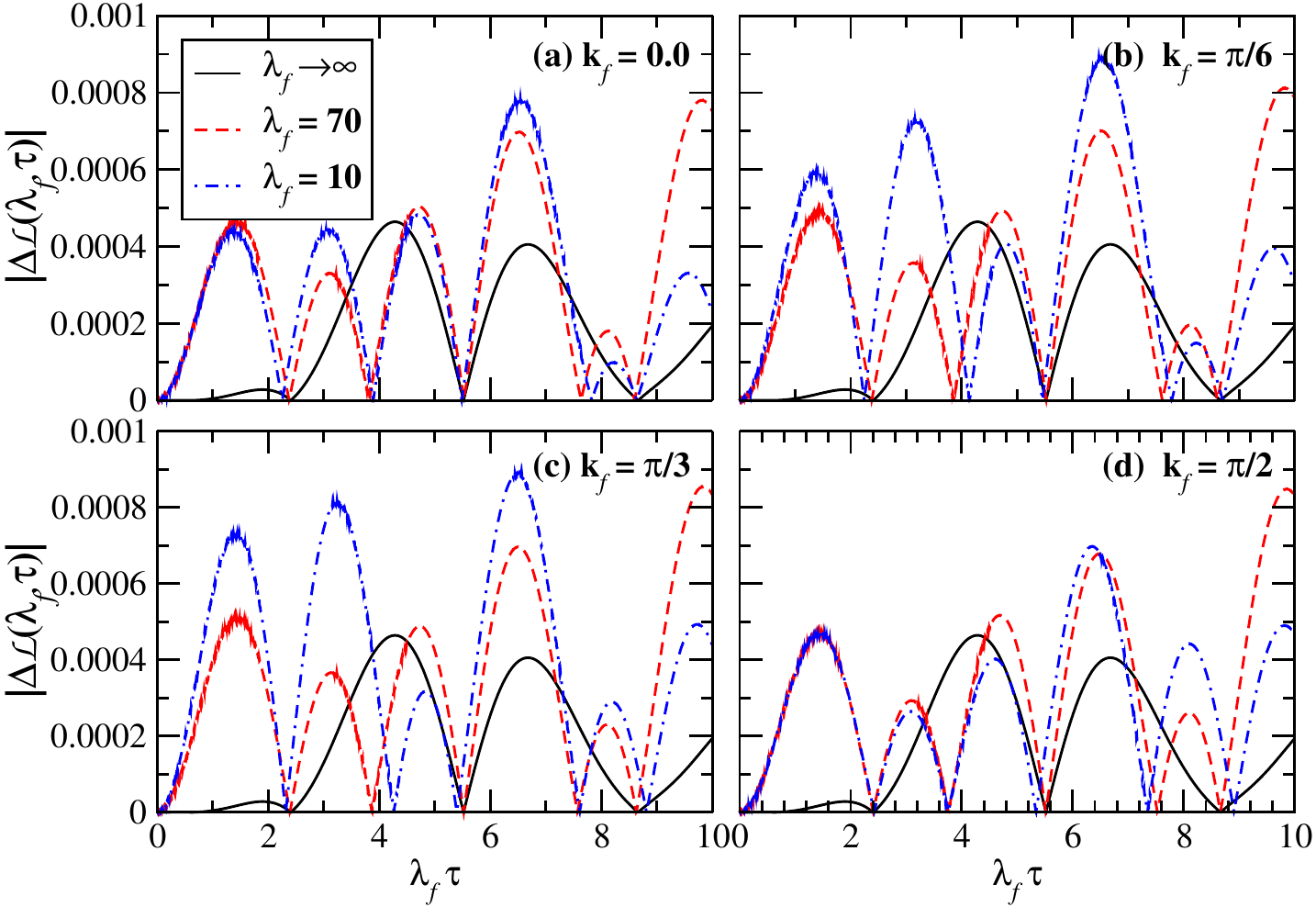}
\par\end{centering}
\caption{(Color online) \textbf{Quench dynamics of the Aubry-André model:} Absolute error of the Loschmidt echo, $\left|\Delta\mathcal{L}(\lambda_{f},\tau)\right|$, for an initially plane wave of wave vector (a) $k = 0$ (band edges), (b) $k = \pi/6$, (c) $k = \pi/3$, and (b) $k = \pi/2$ (band center) with different post-quench modulation incommensurate potential for a system of size $N = 1024$ with $M = 1024$ Chebyshev series. \label{fig:RelativeError}}
\end{figure}

For the correlated Anderson model \cite{Pires2019}, the randomness in the potential manifests itself as a long-range spatially correlated random variable with a spectral density, $S(q)\sim 1/q^{\alpha},$ where $\alpha$ is the correlation strength of the spectral density, controlling the roughness of the potential landscapes. The correlated disorder potential, $\varepsilon_{n},$
is given by \cite{Pires2019,Niaz2020,Niaz2022,Niaz2023CJP},
\begin{equation}
\varepsilon_{n}=\mathcal{A}_{\alpha}\sum_{q=1}^{N/2}\frac{1}{q^{\alpha/2}}\cos\left(\frac{2\pi q}{N}n+\phi_{q}\right),\label{eq:rp0}
\end{equation}
where $\mathcal{A}_{\alpha}$ is a normalization constant that imposes a unit variance of the local potential $(\sigma_{\varepsilon}^{2} = 1)$ with zero average, and $\phi_{q}$ is the $N/2$ independent disorder phases distributed uniformly in
the interval {[}$0,\,2\pi]$. It is worth mentioning that, in the limit of the infinite correlation of the disorder potential, the disorder distribution takes the following sinusoidal form (normalized) of wavelength $N$ with vanishing noise:
\begin{align}
\varepsilon_{n} & =\sqrt{2}\cos\left(\frac{2\pi}{N}n+\phi_{1}\right),\label{eq:SinusoidalForm}
\end{align}
In this limit, the local value of the disorder potential is dominated by the term $q=1$, and the disorder is a static sinusoidal potential with a random phase. As a consequence, the system exhibits metallic behavior due to the (effective) absence of disorder. In this sense, the spectral function of the correlated Anderson model shows similar behavior to the density of state in position space \cite{Niaz2019}. In the limit of $\alpha=0$, the system is insulating in nature with all the eigenstates localized.
\subsection{Loschmidt echo}
\begin{figure}
\begin{centering}
\includegraphics[scale=0.36]{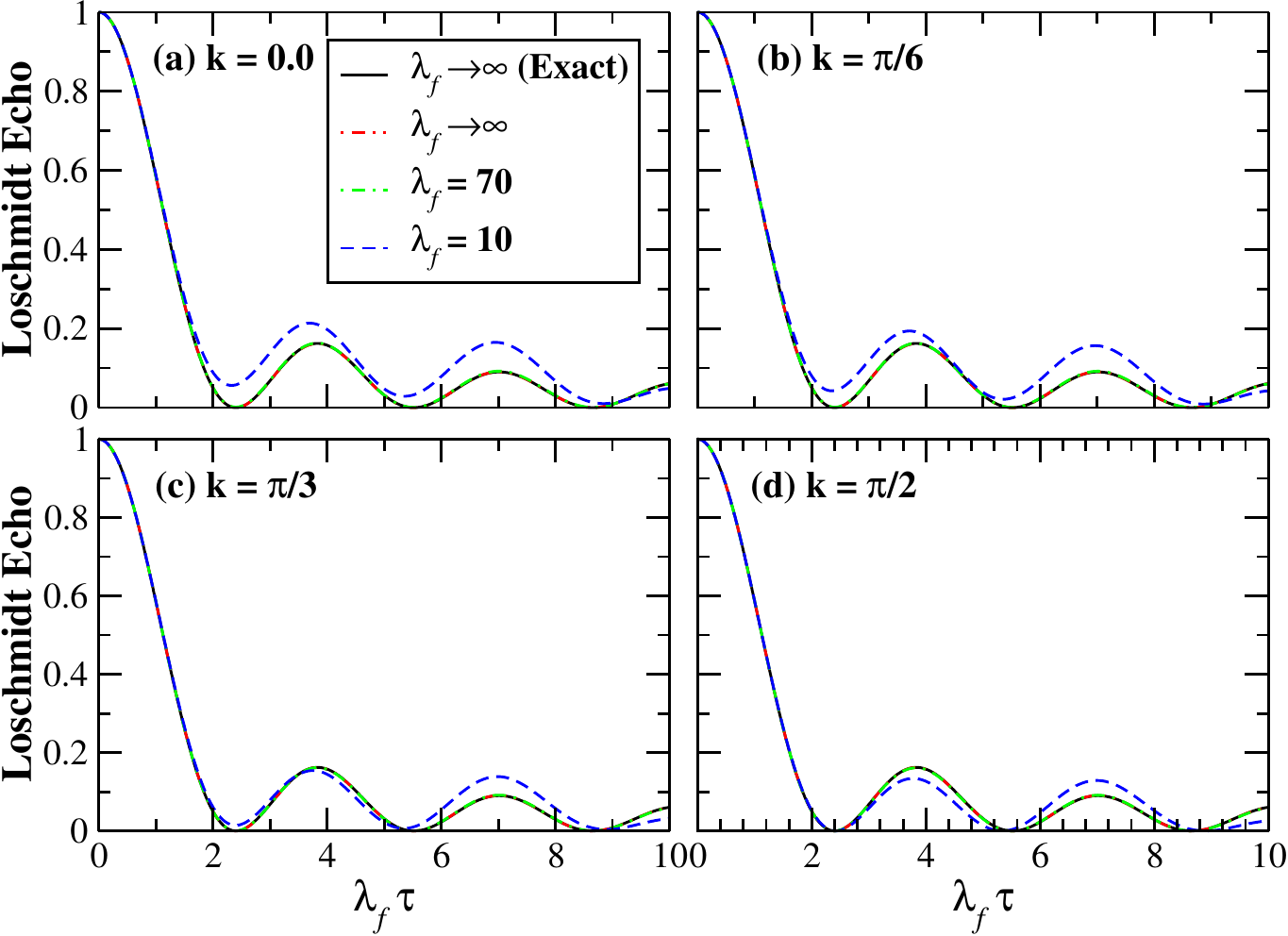}
\par\end{centering}
\caption{(Color online) \textbf{Quench dynamics of the Aubry-André model:} The KPM estimates of the Loschmidt echo for an initially plane wave of wave vector (a) $k = 0$ (band edges), (b) $k = \pi/6$, (c) $k = \pi/3$  and (d) $k = \pi/2$ (band center) with different post-quench modulation incommensurate potential for a system of size $N = 1024$ with $M = 1024$ Chebyshev series. The numerical data are in perfect agreement with the analytical result (black bold
curve) in the limit $\lambda_{f}\rightarrow \infty$.\label{fig:aaLEvst-Ei0-Ef-2p}}
\end{figure}

A Loschmidt echo is the measure of the overlap between an initial reference and its time-evolved state of the system. It quantifies the sensitivity of the non-equilibrium dynamics of quenched systems. A quantum quenching process is an abrupt change in the Hamiltonian $\mathcal{\mathcal{H}}(x)$
of a system, where $x$ is the strength of the prequench parameter. At time $\tau=0$, $\ket{\Psi(x)}$ is the initial ground state of the system with normalization condition, $\braket{\Psi(x)}{\Psi(x)}=1$.  The Hamiltonian $\mathcal{\mathcal{H}}(y)$ governs the time evolution of the system at certain times $\tau>0$, reaching the unitary
evolving state \cite{Yang2017,Yin2018,Xu2021}
\begin{equation}
\ket{\Psi(x,y,\tau)}=e^{-i\tau\mathcal{\mathcal{H}}(y)}\ket{\Psi(x)},
\end{equation}
where $y$ denotes the strength of the postquench parameter. Phenomenologically, the quantum quenches trigger a time-evolving state $\ket{\Psi(x,y,\tau)}$ under the postquench Hamiltonian $\mathcal{\mathcal{H}}(y)$ from a reference state $\ket{\Psi(x)}$. The Loschmidt amplitude, $\mathcal{G}(x,y,\tau)$, of the quenched
system, has the form,
\begin{equation}
\mathcal{G}(x,y,\tau)=\braket{\Psi(x)}{\Psi(x,y,\tau)}.\label{eq:LoschmidtAmplitude}
\end{equation}
A Loschmidt echo, $\mathcal{L}(x,y,\tau)$, is the dynamical version of the ground state fidelity (return probability), defined as \cite{Yang2017,Yin2018,Xu2021},
\begin{equation}
\mathcal{L}(x,y,\tau) \equiv \left|\mathcal{G}(x,y,\tau)\right|^{2} =\left|\braket{\Psi(x)}{\Psi(x,y,\tau)}\right|^{2}.\label{eq:LoschmidtEcho}
\end{equation}
The Loschmidt echo typically decays from unity, oscillating at the same frequency and damping amplitude after some time interval as depicted in Fig.~\ref{fig:aaLEvst-Ei0-Ef-2p} of Ref. \cite{Yang2017}. More importantly, the Loschmidt echo periodically approaches zero at critical times under certain conditions. The occurrence of zeros characterizes the DQPTs.
%
\begin{figure}
\begin{centering}
\includegraphics[scale=0.38]{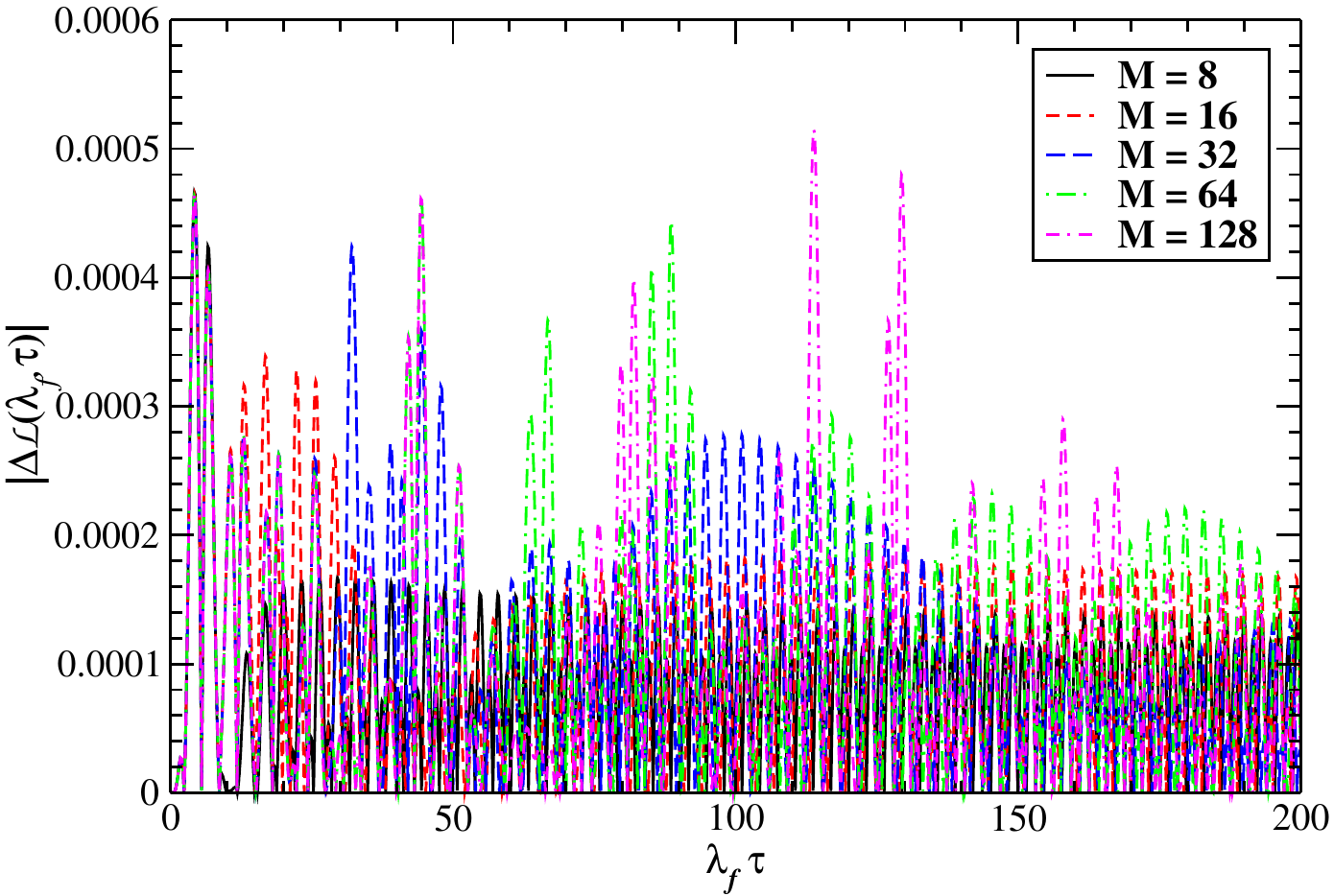}
\par\end{centering}
\caption{(Color online) \textbf{Quench dynamics of the Aubry-André model:} Absolute error of the Loschmidt echo, $\left|\Delta\mathcal{L}(\lambda_{f},\tau)\right|$, for an initially plane wave of wave vector $k = 0$ quenched into a strongly localized time-dependent state with $\lambda_{f}=10^4$. Numerical calculations are carried out for a system of size $N = 1024$ with different degree $M$ of the expansion.
\label{fig:moments}}
\end{figure}
\section{Kernel Polynomial Method\label{sec:KernelPolynomialMethod}}

The KPM \cite{Weibe2006,Simao2020,Niaz2021,Simao2022,BJELCIC2022108477,Castro2023,Li2023,FAN2021,Sobczyk2022}
is an efficient and accurate numerical algorithm that plays a prominent role in condensed matter problems \cite{Simao2020,BJELCIC2022108477,Simao2022}.
It is a method based on polynomial-expansions that yields results of guaranteed accuracy at a low computational cost. Importantly, a Chebyshev polynomial --- with good convergence properties of the target series and a close relation to the Fourier transform \cite{Mason2002}--- turns out to be a reasonable choice of KPM technique for the condensed matter problems.
Furthermore, the accuracy and numerical convergence of the KPM technique can be controlled by the polynomial moments and the optimal kernel.

For the simulations based on the polynomial-expansion, the Hamiltonian and all the associated energy scales must be normalized~\footnote{The Hamiltonian and all energy scales can be normalized by dividing $2Dt+\mathcal{F}$, where $D$ is the dimension of the system, $t$ is the hopping integral, and $\mathcal{F}$ is a number that imposes the Hamiltonian spectrum to be in the interval $\left[-1,\,1\right]$. } in the standard range of orthogonality of the Chebyshev polynomials $(\left[-1,\,1\right])$. Furthermore, the accuracy and numerical convergence of the KPM technique strongly depends on the Gibbs damping factor and the coefficients of the Chebyshev polynomials. The first type of $m^\text{th}$ degree Chebyshev polynomials, $T_{m}(z)$ are defined as follows:
\begin{equation}
T_{m}(z)=\cos(m\,\arccos(z)),\quad m\in\mathbb{N}.\label{eq:polynomial}
\end{equation}
Moreover, the $T_{m}(z)$ obeys the following recurrence relation,
\begin{equation}
T_{m}(z)=2zT_{m-1}(z)-T_{m-2}(z),\,\,\,\,\,\,\,m>1,\label{eq:recurrsion}
\end{equation}
starting with $T_{0}(z)=1$ and $T_{1}(z)=z$, and also satisfying the orthogonality relation,
\begin{align}
\braket{T_{m}(z)}{T_{n}(z)} & =\frac{1}{\pi}\int_{-1}^{1}\,T_{m}(z)T_{n}(z)(1-z^{2})^{-1/2}dz,\nonumber \\
& =\frac{1}{2}\delta_{m,n}(\delta_{m,0}+1).\label{eq:Orthogonality}
\end{align}
For the KPM expansion of Loschmidt amplitude (see Eq.~(\ref{eq:LoschmidtAmplitude})), we make use of the identity \cite{Iacopi2016},
\begin{equation}
e^{-iz\tau}=\sum_{m=0}^{\infty}\frac{2i^{-m}}{1+\delta_{m,0}}J_{m}(\tau)T_{m}(z),\quad|z|\leq1,
\end{equation}
for the $e^{-i\mathcal{H}\tau}$ part of Loschmidt amplitude. Here, $J_{m}(z)$ is the Bessel function of $m^\text{th}$ order. The polynomial-expansion of $e^{-i\mathcal{H}\tau}$ becomes
\begin{equation}
e^{-i\tilde{\mathcal{H}} \Omega\tau}=\sum_{m=0}^{\infty}\frac{2i^{-m}}{1+\delta_{m,0}}J_{m}(\Omega\tau)T_{m}(\tilde{\mathcal{H}}),
\end{equation}
where $\tilde{\mathcal{H}}=\mathcal{H}/\Omega$ is the rescaled Hamiltonian of the system and $\Omega$ is a positive energy scale that normalizes the Hamiltonian to place its spectrum within $\left[-1\ ,1\right]$. The KPM approximated Loschmidt amplitude, $\tilde{\mathcal{G}}(x,y,\tau)$, has the following form:
\begin{align}
\tilde{\mathcal{G}}(x,y,\tau) & =\sum_{m=0}^{\infty}\frac{2i^{-m}}{1+\delta_{m,0}}J_{m}(\Omega\tau)\bra{\Psi(x)}T_{m}(\mathcal{\tilde{H}}(y))\ket{\Psi(x)},\label{eq:Green2}
\end{align}
In practical numerical calculations, the KPM approximations of a target function can only be carried out for a finite Chebyshev series. Therefore, the truncated form of the Loschmidt amplitude is,
\begin{equation}
\tilde{\mathcal{G}}(x,y,\tau)=\sum_{m=0}^{M-1}\frac{2i^{-m}}{1+\delta_{m,0}}J_{m}(\Omega\tau)\average{T_{m}(\mathcal{\tilde{H}}(y))},\label{eq:GLEchoKPM}
\end{equation}
\begin{figure}
\begin{centering}
\includegraphics[scale=0.36]{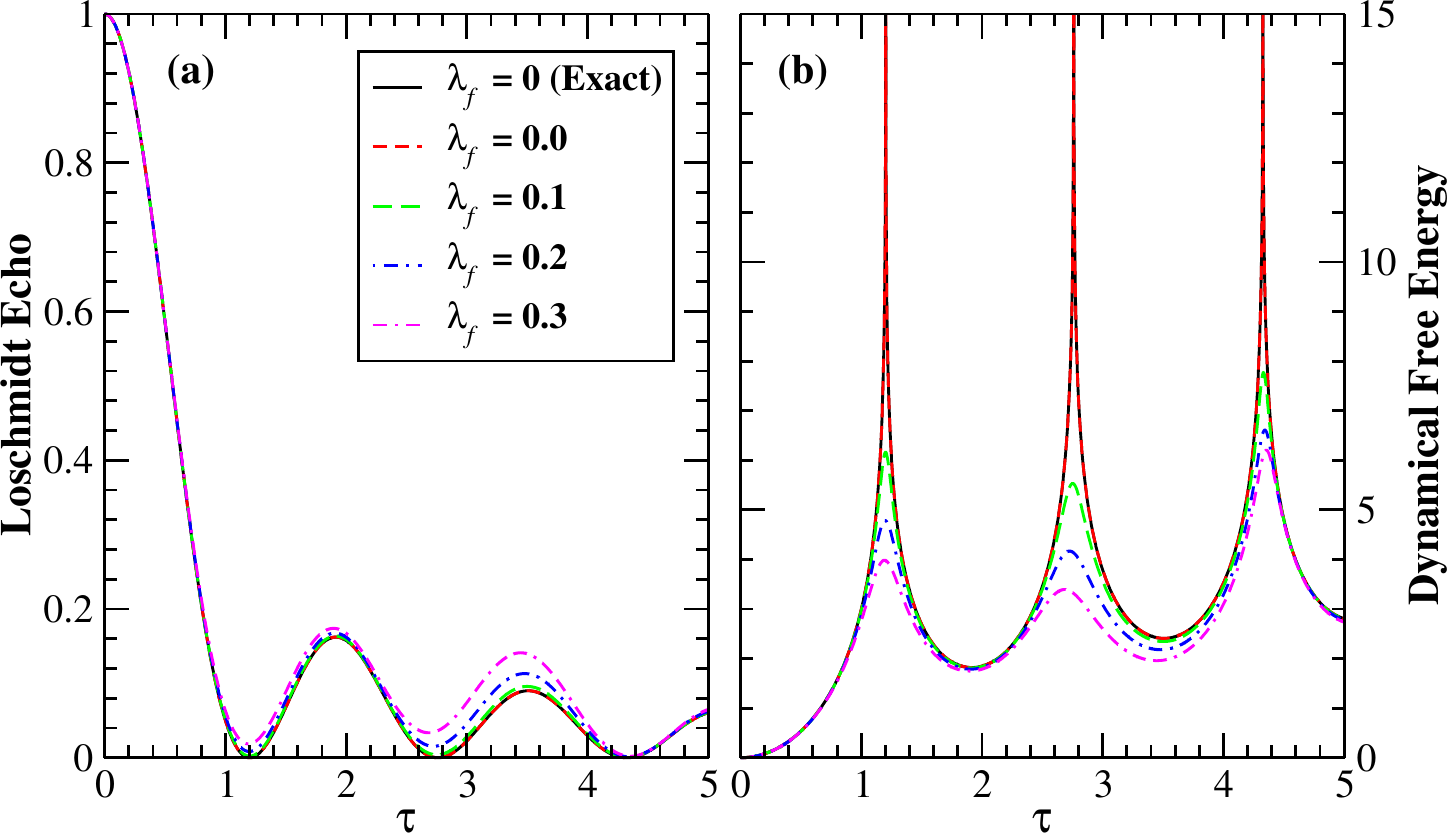}
\par\end{centering}
\caption{(Color online) \textbf{Quench dynamics of the Aubry-André model:} (a) The Chebyshev polynomial simulations of the Loschmidt echo for an initially localized state $(\lambda_{i}=1000)$ with various small postquench modulation incommensurate potentials for a system of size $N = 1024$ with $M = 1024$ Chebyshev series. (b) Evolution of the dynamic free energy of the quantum quench system. The numerical data are in perfect agreement with the analytical calculation (black bold curve) at $\lambda_{f}=0$.\label{fig:aaLEvst-Ei-Ef0-2p}}
\end{figure}
%
where
\begin{align}
\average{T_{m}(\mathcal{\tilde{H}}(y))} & =\bra{\Psi(x)}T_{m}(\mathcal{\tilde{H}}(y))\ket{\Psi(x)},\nonumber \\
& =\braket{\Psi(x)}{\Psi_{m}(x)},\label{eq:ExpectationValue}
\end{align}
is the expectation value of the Chebyshev polynomials in the Hamiltonian and $\ket{\Psi_{m}(x)}= T_{m}(\mathcal{\tilde{H}}(y)) \ket{\Psi(x)}$ are the $m^\text{th}$ order polynomial-dependent states. This truncated approximation works well for the Loschmidt echo. However, for non-differentiable functions, the truncated Chebyshev expansion leads to unwanted oscillations known as Gibbs oscillations, which can be eliminated by using an optimized damping factor \cite{Weibe2006}.

For the polynomial approximation of the Loschmidt echo, we start with the initial vector $\ket{\Psi(x)}$ at all instants of discrete time $\tau$. The main focus is to implement KPM for Loschmidt echo calculation when initial eigenstates $(\ket{\Psi(x)})$ are extended or localized. The initial eigenstates of the system Hamiltonian (Eq.~(\ref{eq:1DHamiltonian})) in the absence of the local potential $(\varepsilon_{n}=0)$ are plane wave, $\ket{k}$,
\begin{equation}
\ket{\Psi_{0}(\varepsilon_{n}=0)}\equiv\ket{k}=\frac{1}{\sqrt{N}}\sum_{n=1}^{N}\exp\left(ikn\right)\hat{c}_{n}^{\dagger}\ket{0},\label{eq:Planewave}
\end{equation}
where $k$ is the wave vector lying in the first Brillouin zone, \textit{i.e.,} $k\in \left(-\pi/a\ ,\pi/a\right]$ with a lattice spacing $a$. The corresponding eigenvalue is $E = 2t\cos(ka)$. On the other hand, the initial Hamiltonian state of the system Hamiltonian (Eq.~(\ref{eq:1DHamiltonian})) in the presence of an infinite diagonal potential $(\varepsilon_{n}\rightarrow\infty)$ is localized at a single site $s$,
\begin{equation}
\ket{\Psi_{s}(\varepsilon_{n}\rightarrow\infty)}=\sum_{n=1}^{N}\delta_{ns}\hat{c}_{n}^{\dagger}\ket{0}.\label{eq:Localizedwave}
\end{equation}
Importantly, the expectation value of Chebyshev polynomials in the Hamiltonian (see Eq.~(\ref{eq:ExpectationValue})) is independent of the time scale, which can be handled straightforwardly using the recursion relations for $T_{m}(\mathcal{\tilde{H}}(y))$ (Eq.~(\ref{eq:recurrsion})). For instance, starting from an initial state $\ket{\Psi(x))}=\ket{\Psi_{0}(x)}$, one can iteratively construct the states $\ket{\Psi_{m}(x)}= T_{m}(\mathcal{\tilde{H}}(y)) \ket{\Psi_{0}(x)}$ for the expectation values of the $m^{th}$ degree polynomial $T_{m}(\mathcal{\tilde{H}}(y))$ as follows:
\begin{align}
\ket{\Psi_{0}(x)} & =T_{0}(\mathcal{\tilde{H}}(y))\ket{\Psi(x))}, \\
\ket{\Psi_{1}(x)} & =T_{1}(\mathcal{\tilde{H}}(y)) \ket{\Psi_{0}(x)}=\mathcal{\tilde{H}}(y)\ket{\Psi_{0}(x)},\label{eq:RecurenceState}
\end{align}
and using the recursion relation
\begin{equation}
\ket{\Psi_{m}} = 2\mathcal{\tilde{H}}(y)\ket{\Psi_{m-1}} - \ket{\Psi_{m-2}}.\label{eq:RecurenceStateS2}
\end{equation}
The expectation value of the $T_{m}(\mathcal{\tilde{H}}(y))$ turns out,
\begin{align}
\average{T_{0}(\mathcal{\tilde{H}}(y))} & =\braket{\Psi_{0}}{T_{0}(\mathcal{\tilde{H}}(y))|\Psi_{0}}=1, \\
\average{T_{1}(\mathcal{\tilde{H}}(y))} & =\braket{\Psi_{0}}{\mathcal{\tilde{H}}(y)|\Psi_{0}}=\average{\mathcal{\tilde{H}}(y)},\label{eq:T0T1}
\end{align}
and for $m>1$,
\begin{equation}
\average{T_{m}(\mathcal{\tilde{H}}(y))}=2\average{\mathcal{\tilde{H}}(y)} \average{T_{m-1}(\mathcal{\tilde{H}}(y))}-\average{T_{m-2}(\mathcal{\tilde{H}}(y))}.
\end{equation}
It is noted that the expectation value of the Chebyshev polynomial $T_{m}(\mathcal{\tilde{H}}(y))$ is independent of time; therefore, we may not need to compute it again for various time steps.

The computational complexity of the KPM estimates of Loschmidt echo is $\mathcal{O}(SMN)$ for a sparse matrix of the system Hamiltonian, where $S$ is the sample average over realizations of disorder (for disordered systems), and $M$ is the number of Chebyshev series. Iterative computation of the expectation value $\average{T_{m}(\mathcal{\tilde{H}}(y))}$ is the most time-consuming part of the expansion approach, costing $\mathcal{O}(N)$ numerical complexity and determining the performance of the KPM. The $\mathcal{O}(SM)$ comes from averaging over disorder realizations and summing over the Chebyshev series in Eq.~(\ref{eq:GLEchoKPM}). However, for a dense Hamiltonian matrix, the computational cost of KPM becomes $\mathcal{O}(SMN^{2})$ due to multiplications for all elements of $\mathcal{\tilde{H}}(y)$ and the initial state $\ket{\Psi(x)}$. The computational cost is very effective compared to the EDM, which costs $\mathcal{O}(N^{3})$, can be found in Appendix~(\ref{appendix:appendixA}). Furthermore, the numerical convergence and resolutions of the KPM estimates of the Loschmidt echo are controlled by the number of Chebyshev series $(M)$. This means that the absolute difference between the exact Loschmidt echo $\mathcal{L}(x,y,\tau)$ and the KPM approximated Loschmidt echo $\tilde{\mathcal{L}}(x,y,\tau)$ goes to zero.
\begin{equation}
\left|\mathcal{L}(x,y,\tau)-\tilde{\mathcal{L}}(x,y,\tau)\right|{\rightarrow}0,
\end{equation}
for sufficiently large Chebyshev series in the $N\rightarrow\infty$ limit. The KPM estimates of the Loschmidt amplitude converge uniformly for a sufficiently large Chebyshev series.
%
\begin{figure}
\begin{centering}
\includegraphics[scale=0.36]{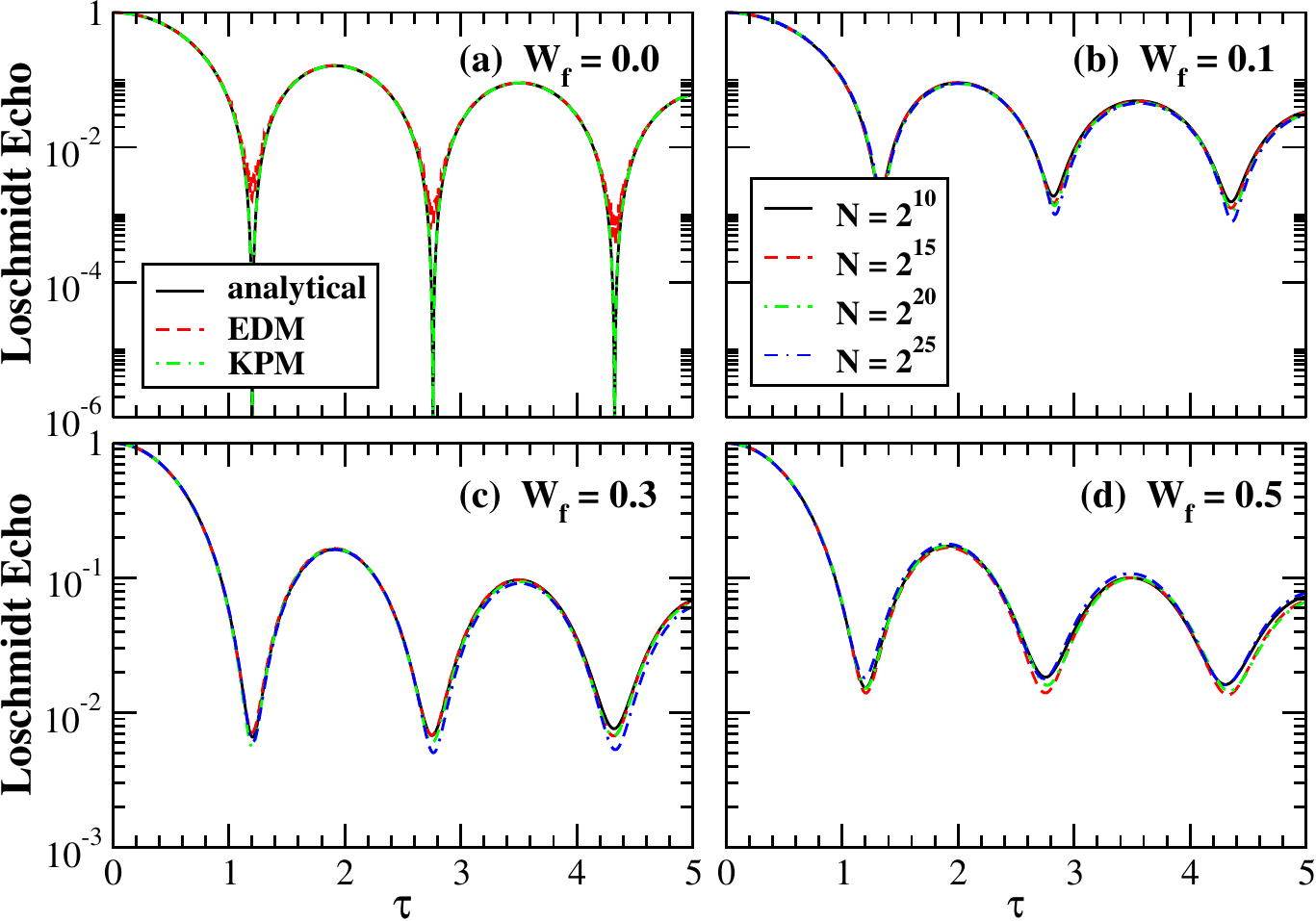}
\par\end{centering}
\caption{(Color online) \textbf{Quench dynamics of the standard Anderson model:} Evolution of the Loschmidt echo in the log-linear scale for initially localized eigenstates of the Hamiltonian with $(W_{i} \rightarrow \infty)$ quenched into a time-evolved state with postquench disorder strengths (a) $W_{f}=0.0$, (b) $W_{f}=0.1$, (c) $W_{f}=0.3$, and (d) $W_{f}=0.5$. For $W_{f}=0.0$, the numerical calculations (KPM and EDM) are carried out for a system of size $N = 1024$ with $M = 1024$ Chebyshev moments and compared with the analytical result. \label{fig:wLE-WiInf-Wf0}}
\end{figure}
\section{Applications of the Kernel Polynomial Method\label{sec:ApplicationsKPM}}
In this section, we demonstrate practical applications of the polynomial-expansion of the Loschmidt echo for quenched electronic systems. We cover the quench dynamics of lattice models with infinitely large system sizes, leading to a qualitative improvement in understanding the model.
\subsection{Aubry--André model}\label{subsec:AAModel}
Our main goal is to efficiently compute the polynomial-expansion of the Loschmidt echo of a quenched quantum system. We validate our KPM approach on a well-established Aubry--André model \cite{Yang2017}. In what follows, we demonstrate the polynomial estimates of the Loschmidt echo of a 1D Aubry--André model where quench dynamics are induced by an abrupt change in the strength of the incommensurate potential. To get a better understanding, we first consider two limiting cases of quench processes,
\begin{itemize}
\item The initial plane wave state (Eq.~\ref{eq:Planewave}) of the system Hamiltonian with $\lambda_{i}=0$ is quenched into a strongly localized time-evolved state with $\lambda_{f} \rightarrow \infty$.
\item The initial localized state (Eq.~\ref{eq:Localizedwave}) of the system Hamiltonian with $\lambda_{i}\rightarrow \infty$ is quenched into an time-evolved extended state with $\lambda_{f}=0$.
\end{itemize}
Here, $\lambda_{i}$ and $\lambda_{f}$ are the prequench and postquench incommensurate potential controlling parameters, respectively. In the former case  $(\lambda_{i}=0,\quad\lambda_{f}\rightarrow \infty)$, one can analytically calculate the Loschmidt echo of the system in the thermodynamic limit,
\begin{align}
\mathcal{L}(\lambda_{i}=0,\lambda_{f}\rightarrow \infty,\tau) & =\left|J_{0}(\lambda_{f}\tau)\right|^{2},\label{eq:Exact-ei0-efInf}
\end{align}
where $J_{0}(x_{s})$ is the zero-order Bessel function of the $1^{st}$ kind, of a series of zeros $x_{s}$ with $s\in\mathbb{N}$ set of positive roots. In the latter case $(\lambda_{i} \rightarrow \infty,\quad\lambda_{f}=0)$, one can obtain straightforwardly,
\begin{align}
\mathcal{L}(\lambda_{i} \rightarrow \infty,\lambda_{f}=0,\tau) & =\left|J_{0}(2\tau)\right|^{2},\label{eq:Exact-eiInf-ef0}
\end{align}
in the thermodynamic limit. An absolute error of the Loschmidt echo $|\Delta \mathcal{L}(\lambda_{f},\tau)| = \left|\mathcal{L}(\lambda_{f},\tau)-\tilde{\mathcal{L}}(\lambda_{f},\tau)\right| $ for an initial plane wave of the Aubry--André model is illustrated in Fig.~\ref{fig:RelativeError}. It is important to mention that $\mathcal{L}(\lambda_{f},\tau)$ is calculated analytically for $\lambda_{f}\rightarrow \infty$ in the thermodynamic limit. On the other hand, for finite $\lambda_{f}$, $\mathcal{L}(\lambda_{f},\tau)$ is calculated numerically by EDM. For various $\lambda_{f}$, the KPM technique estimates the Loschmidt echo up to 4 decimal points. It is noted that for a large enough system's size, the accuracy and numerical convergence of the polynomial-expansion scheme depend on the Chebyshev moments. Most importantly, the high accuracy of the scheme can be traced to the fact that when $m>\tau$, $J_{m}(\tau)$ exponentially goes to zero. Thus, the degree of Chebyshev moments has to be at least $\tau$ for better accuracy. Figure~\ref{fig:moments} demonstrates the role of the degree of Chebyshev moments for a fixed system size at large time $\tau$ for wave vector $k=0$. An absolute error in Loschmidt echo is calculated for a system of size $N=1024$ with $\lambda_{i}=0$ and $\lambda_{f}=10^4$. It is clear that the KPM estimates show excellent agreement with the analytical result for large enough Chebyshev moments up to $4$ decimal points at large time. However, there are a number of drawbacks to Chebychev polynomial-expansion technique. First, the Chebychev-expansion scheme cannot propagate wave packets of a time-dependent Hamiltonian. Another drawback is that the polynomial-expansion scheme fails to reproduce the long time duration of propagation where the order of the Bessel function is smaller than the time.

The KPM estimates of Loschmidt echo under the Aubry--André model for the initial plane wave state quenched into a time-evolved localized state are shown in Fig.~\ref{fig:aaLEvst-Ei0-Ef-2p}. All numerical calculations are performed for the system of size $N=1024$ with various postquench modulation potentials at $\lambda_{i}=0$. The polynomial approximations of the Loschmidt echo for Chebyshev moments $M=1024$ have a series of zeros at critical times, as reported in the literature \cite{Yang2017}. We also investigate the role of the wave vector $(k)$ on the quench dynamics of the system. For $k=0$, the system displays DQPTs for larger $\lambda_{f}$, characterized by the disappearing value of the Loschmidt echo in the thermodynamic limit, as shown in Fig.~\ref{fig:aaLEvst-Ei0-Ef-2p}(a). On the other hand, for $0<k \leq \pi/2$, the quench dynamics show a clear signature of the DQPTs even for $\lambda_{f}=10$ as depicted in Fig.~\ref{fig:aaLEvst-Ei0-Ef-2p}(b-d). The quench dynamics turn out to be wave vector independent and in excellent agreement with the analytical thermodynamic results (Eq.~(\ref{eq:Exact-ei0-efInf})) in the limit of infinite postquench modulation potential $(\lambda_{f} \rightarrow \infty)$. It is a well-established fact that the Aubry--André model is self-dual under the Fourier transformation at the critical quasiperiodic potential $\lambda_{c} = 2$, leading to a quantum phase transition without mobility edges \cite{Sarma1988,Biddle2010,Ganeshan2015}. Analogously, the quench dynamics turn out to be energy-independent, reflecting wave vector-independent DQPTs of the model.
\begin{figure}
	\begin{centering}
		\includegraphics[scale=0.36]{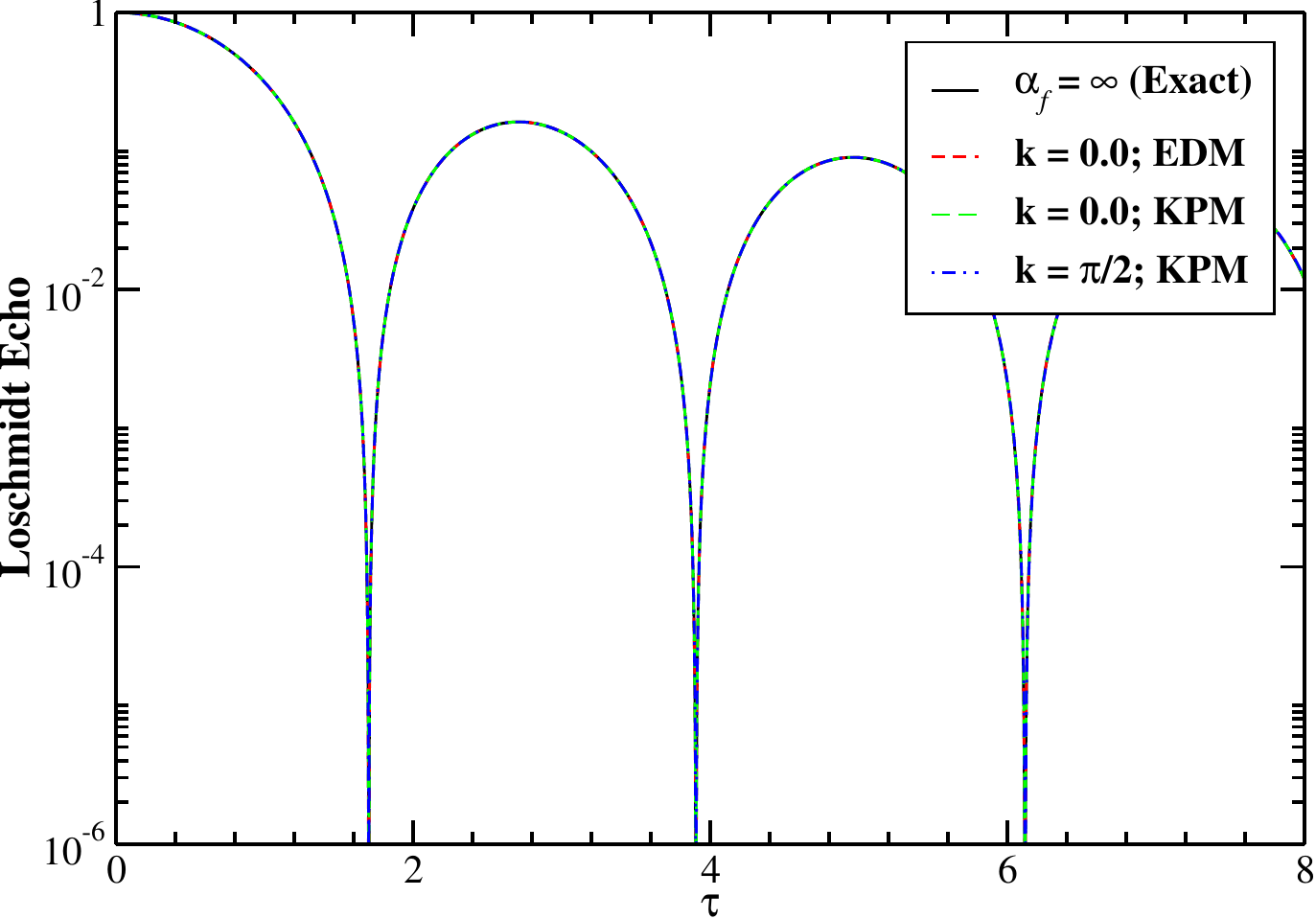}
		\par\end{centering}
	\caption{(Color online) \textbf{Quench dynamics of the correlated Anderson model:} The numerical and analytical calculations of the Loschmidt echo in the log-linear scale for an initially plane wave of wave vector $k = 0$ (band edges) and $k = \pi/2$ (band center) quenched in a strongly correlated disorder regime. Numerical calculations are carried out for a system of size $N = 1024$ with $M = 1024$ Chebyshev series at $\alpha_{f} = 1000$.
		\label{fig:LE-ei0-ef-CorrelationEffect}}
\end{figure}

The nonequilibrium dynamics under prequench localized and postquench extended states based on the polynomial-expansion of the Loschmidt echo are illustrated in Fig.~\ref{fig:aaLEvst-Ei-Ef0-2p}. Similar to the previous case, the system exhibits DQPTs, characterized by the vanishing values of the Loschmidt echo at the critical time, as shown in Fig.~\ref{fig:aaLEvst-Ei-Ef0-2p}(a). All numerical calculations are carried out for a system of size $N=1024$, Chebyshev moments $M=1024$, $\lambda_{i}=1000$, and various postquench modulation potentials. Fig.~\ref{fig:aaLEvst-Ei-Ef0-2p}(b) shows the evolution of the dynamical free energy, $\mathcal{E}(\lambda_{i},\lambda_{f},\tau)$. The black bold curve corresponds to the analytical result,
\begin{align}
\mathcal{E}(\lambda_{i} \rightarrow \infty,\lambda_{f}=0,\tau) & =-\ln\left|J_{0}(2\tau)\right|^{2},
\end{align}
in the thermodynamic limit. For a fixed system size, a small deviation of the estimated Loschmidt echo is observed for small nonzero values of $\lambda_{f}$, which begins to disappear with decreasing $\lambda_{f}$. Furthermore, the occurrence of zeros in the Loschmidt echo corresponding to divergences in the dynamical free energy can be viewed as a sign of the DQPTs.
\subsection{Standard Anderson model}\label{subsec:wAModel}
We now discuss the quench dynamics in the uncorrelated Anderson model, where the quench is characterized by an abrupt change in the strength of the diagonal random potential. We consider the case where an initial localized state of the prequench Hamiltonian with infinite disorder strength is quenched into a time-evolved extended state of the postquench Hamiltonian with $W_{f}=0$. The time-evolved extended eigenstates of the postquench Hamiltonian $\hat{\mathcal{H}}(W_{f})$ are plane-wave states $\ket{\Psi(W_{f})}=\ket k$ with eigenenergy $E_{k}=2t\cos(ka)$. The Loschmidt amplitude (Eq.~(\ref{eq:LoschmidtAmplitude})) becomes,
\begin{equation}
\mathcal{G}(W_{i},\tau)=\expectation{\Psi_{m}(W_{i})}{ e^{-i\tau\hat{\mathcal{H}}(W_{f})}}{\Psi_{m}(W_{i})}, \label{eq:wLAmplitude-WiInf-Wf0-1}
\end{equation}
where $\ket{\Psi_{m}(W_{i})}$ are the eigenstates of the postquench Hamiltonian, localized at a single site $m$. In this case, the Loschmidt echo can be written as:
\begin{align}
\mathcal{G}(W_{i},\tau) & =\expectation{\Psi_{m}(W_{i})}{ e^{-i\tau\hat{\mathcal{H}}(W_{f})}}{k} \braket{k}{\Psi_{m}(W_{i})}, \nonumber \\
& =\frac{1}{N}\sum_{m=1}^{N}e^{-2i\tau\cos(ka)}\left| \braket{\Psi_{m}(W_{i})}{k}\right|^{2},\nonumber \\
& =\frac{1}{N}\sum_{m=1}^{N}e^{-2i\tau\cos(ka)},\label{eq:Wiamplitude}
\end{align}
where the wave vector $k\in(-\pi/a,\pi/a]$. The expression Eq.~(\ref{eq:Wiamplitude}) in the thermodynamic limit becomes,
\begin{align}
\mathcal{G}(\alpha_{f}=\infty,\tau) & =\frac{1}{2\pi}\int_{-\pi}^{\pi}dk e^{-2i\tau\cos(ka)},\nonumber \\
& =J_{0}(2\tau).\label{eq:wLAmplitude-WiInf-Wf0}
\end{align}
The nonequilibrium dynamics under the standard Anderson model for an initially localized state $(W_{i}=\infty)$ quenched into a time-evolved state are displayed in Fig.~\ref{fig:wLE-WiInf-Wf0}. For $W_{f}=0$, the numerical obtained by KPM and EDM for a system of size $N=1024$ with $M=1024$ Chebyshev moments show excellent agreement with the analytical result, $\mathcal{L}(W_{i}=\infty,W_{f}=0,\tau)=\left|J_{0}(2\tau)\right|^{2}$, as shown in Fig.~\ref{fig:wLE-WiInf-Wf0}(a). The existence of the DQPTs decreased by increasing the strength of the postquench modulation disorder displayed in Fig.~\ref{fig:wLE-WiInf-Wf0}(b)--(d). This is because the prequench and postquench disorder parameters remain in the same (localized) regime. One can clearly see the size scaling of the Loschmidt echo. The numerical data begin to converge with the increasing size of the system.
\begin{figure}
\begin{centering}
\includegraphics[scale=0.36]{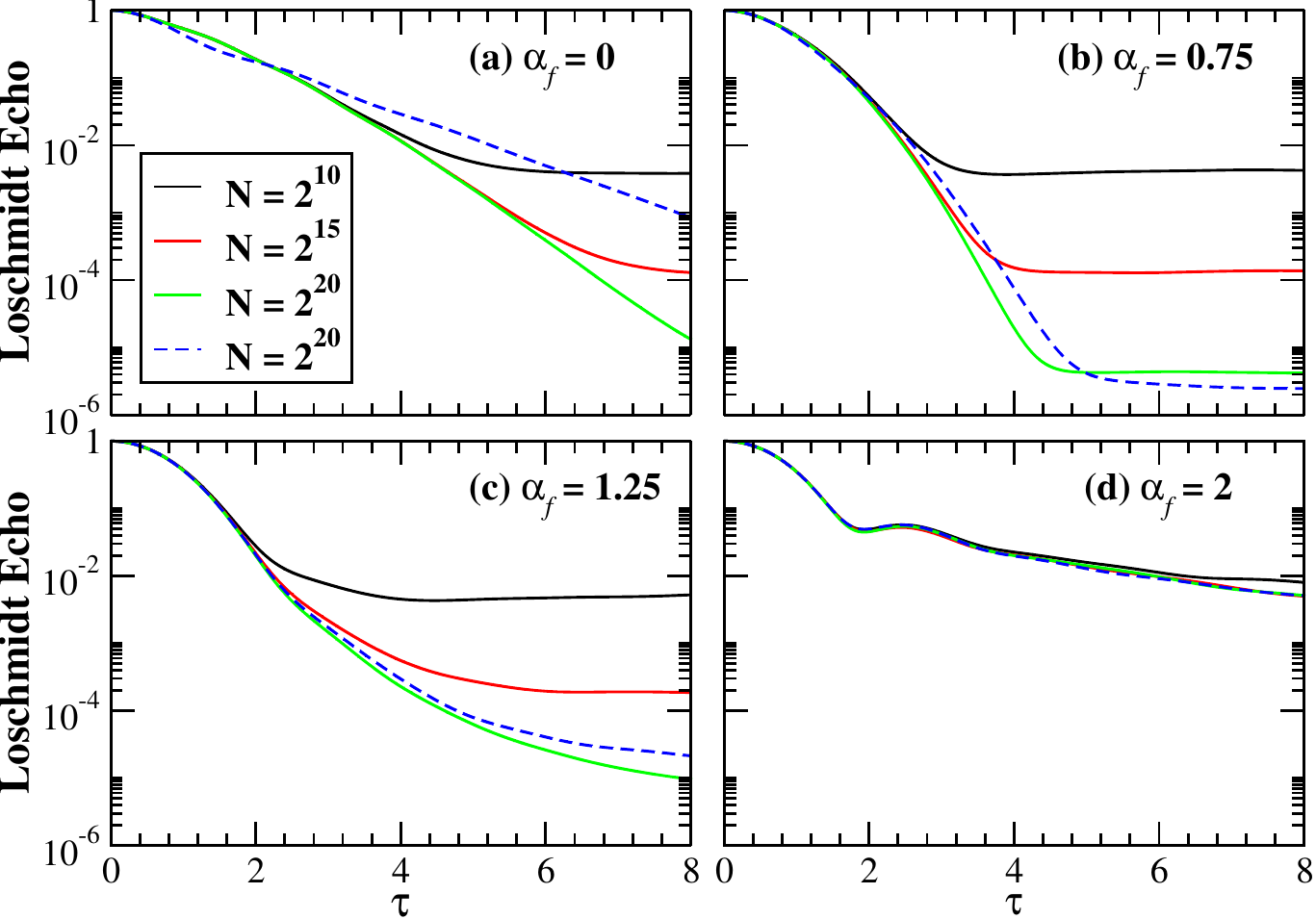}
\par\end{centering}
\caption{(Color online) \textbf{Quench dynamics of the correlated Anderson model:} The Chebyshev polynomial approximation of the Loschmidt echo in log-linear scale for an initial plane wave quenched into a time-evolved correlated disorder state for (a) $\alpha_{f} = 0$, (b) $\alpha_{f} = 0.75$, (c) $\alpha_{f} = 1.25$, and (d) $\alpha_{f} = 2$. Numerical calculations are carried out for different system sizes at wave vectors $k = 0$ (bold curves) and $k = \pi/2$ (dashed curves) with $M = 1024$ Chebyshev series and averaged over $1024$ realizations of disorder.
\label{fig:LE-ei0ef-Sizes-Correlated}}
\end{figure}
%
\subsection{Correlated Anderson model}\label{subsec:cAModel}
We now focus on the quench dynamics of the correlated disordered system, where the quench is characterized by an abrupt change in the strength of spatial correlations of the diagonal random potential. The system is initially in a state $\ket{\Psi(\alpha_{i})}$, which
is the eigenstate of the system Hamiltonian $\hat{\mathcal{H}}(\alpha_{i})$
of the prequenched modulation correlation strength $\alpha_{i}$ at time
$\tau=0$ and $\ket{\Psi(\alpha_{i},\alpha_{f},\tau)}$ be the time-evolving state after performing a sudden quench to the final Hamiltonian $\hat{\mathcal{H}}(\alpha_{f})$.
We define the Loschmidt echo as:
\begin{equation}
\mathcal{L}(\alpha_{i},\alpha_{f},\tau)=\left|\braket{\Psi(\alpha_{i})}{\Psi(\alpha_{i},\alpha_{f},\tau)}\right|^{2},\label{eq:LoschmidtEcho-1}
\end{equation}
where $\alpha_{f}$ defines the strength of the postquench modulation correlation at time $\tau$. The initial eigenstate of the prequench Hamiltonian $\hat{\mathcal{H}}(\alpha_{i})$ is a
plane-wave state $\ket{\Psi(\alpha_{i})}=\ket k$ with eigenenergy,
$E_{k}=2t\cos(ka)$, in the absence of random potential $(\varepsilon(\alpha_{i})=0)$. After performing a sudden quench in the internal correlations of the disorder potential, the Loschmidt amplitude (Eq.~(\ref{eq:LoschmidtAmplitude})) becomes,
\begin{equation}
\mathcal{G}(\alpha_{f},\tau)=\braket{k|e^{-i\tau\hat{\mathcal{H}}(\alpha_{f})}}k.\label{eq:LAmplitude-ei0}
\end{equation}
When an initial extended state is quenched into a strongly correlated
regime $(\alpha_{f}=\infty)$. Then, all the eigenstates $\ket{\Psi_{m}(\alpha_{f})}$
of the postquench Hamiltonian are delocalized with eigenenergy $E_{m}=\sqrt{2}\cos\left(\frac{2\pi}{N}m+\phi_{1}\right)$.
In this case, the Loschmidt echo can be represented as:
\begin{align}
\mathcal{G}(\alpha_{f}=\infty,\tau) & =\sum_{m=1}^{N}\bra ke^{-i\tau\hat{\mathcal{H}}(\alpha_{f})}\ket{\Psi_{m}(\alpha_{f})}\braket{\Psi_{m}(\alpha_{f})}k,\nonumber \\
& =\frac{1}{N}\sum_{m=1}^{N}e^{-i\sqrt{2}\tau\cos\left(\frac{2\pi}{N}m+\phi_{1}\right)},\nonumber \\
& =\frac{1}{N}\sum_{m=1}^{N}e^{-i\sqrt{2}\tau\cos(\varphi)},\label{eq:amplitude}
\end{align}
where $\varphi=(\frac{2\pi}{N}m+\phi_{1})$ is the phase, randomly distributed between $-\pi$ and $\pi$ in the thermodynamic limit. In this limit, the expression Eq.~(\ref{eq:amplitude}) becomes,
\begin{align}
\mathcal{G}(\alpha_{f}=\infty,\tau) & =\frac{1}{2\pi}\int_{-\pi}^{\pi}d\varphi e^{-i\sqrt{2}\tau\cos\left(\varphi\right)},\nonumber \\
& =J_{0}(\sqrt{2}\tau).\label{eq:LAmplitude-ei0-aInf}
\end{align}
The nonequilibrium dynamics under the correlated Anderson model for an initial extended state $(\varepsilon(\alpha_{i})=0)$ quenched into a strongly disordered time-evolved state is illustrated in Fig.~\ref{fig:LE-ei0-ef-CorrelationEffect}. The KPM approximations are carried out for the system of size $N=1024$ with $M=1024$ Chebyshev moments. We obtain an excellent agreement between the analytical expression of Loschmidt echo, $\mathcal{L}(\alpha_{f}=\infty,\tau)=\left|J_{0}(\sqrt{2}\tau)\right|^{2}$, and the numerical results. Importantly, the Loschmidt echo exhibits singularities in time scale, independent of the wave vector, indicating DQPTs induced by correlations in the disorder potential. Fig.~\ref{fig:LE-ei0ef-Sizes-Correlated} illustrates the role of spatial correlations in the disorder potential on the quench dynamics of the model. The initial state is fixed to be the ground state of the prequenched Hamiltonian with zero potential. The KPM estimates of the Loschmidt echo are carried out for $k=0$ and $k=\pi/2$ for different system sizes and $M=1024$ Chebyshev moments with $1024$ realization of disorder. The Loschmidt echo strongly depends on the wave vector in the weakly correlated regime $(\alpha_{f}\lesssim1)$, slowly decaying with the evolving time for $k=\pi/2$. However, the quench dynamics become wave-vector independent in the limit of strong disorder correlations.
\begin{figure}
\begin{centering}
\includegraphics[scale=0.35]{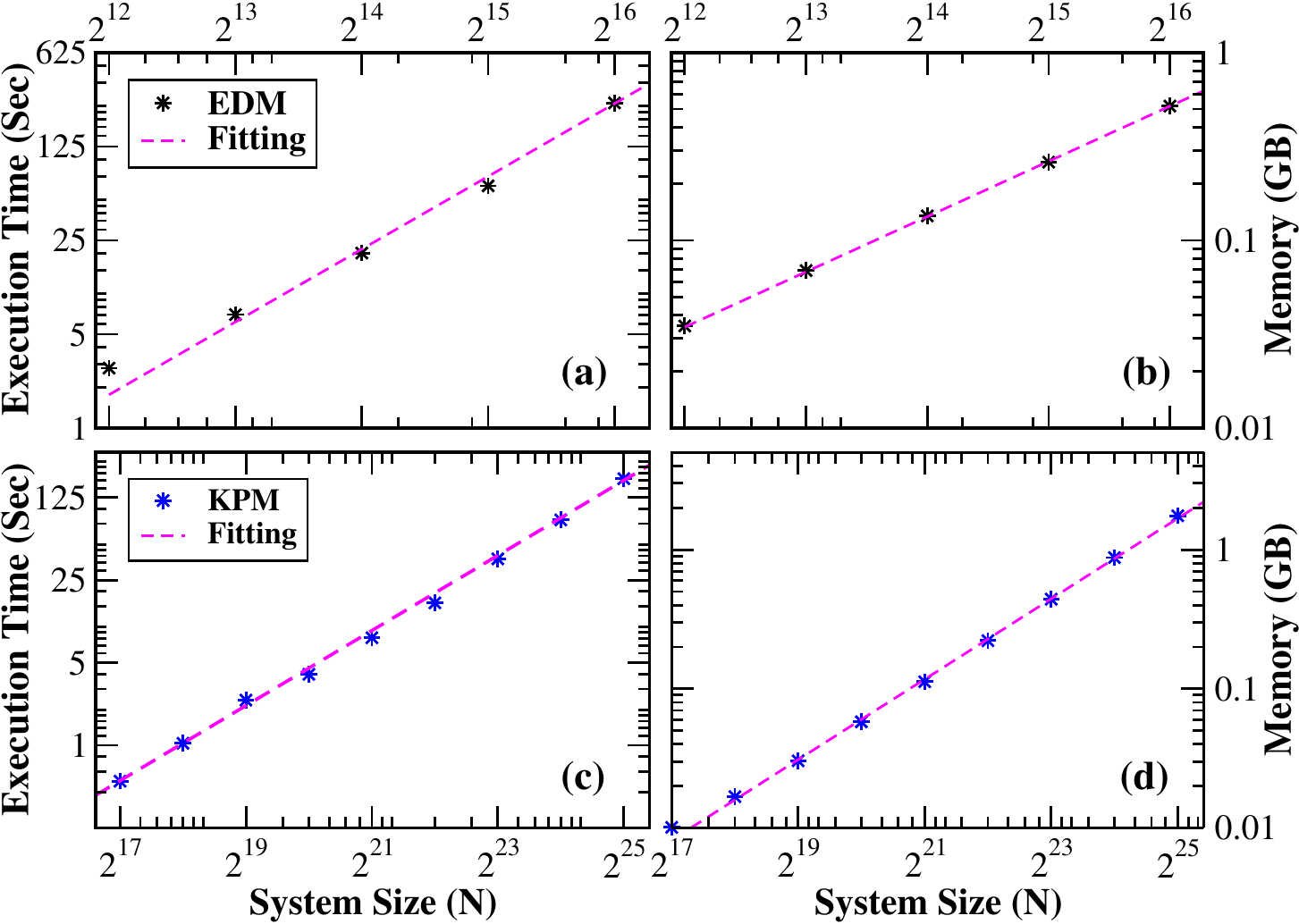}
\par\end{centering}
\caption{(Color online) Scaling of the (a,c) execution time (sec) and (b,d) memory usage (GB) for calculating the Loschmidt echo of the Anderson model based on the exact diagonalization (upper panels) and the polynomial-expansion technique (lower panels). Numerical computations are performed for an initial plane wave ($k=0$) quenched to a time-evolved state with $W_{f}=10$ for the standard Anderson model with $M = 1024$ Chebyshev moments and a single realization of disorder. The magenta dashed curves are the corresponding fits to the numerical data.\label{fig:Complexity}}
\end{figure}
\section{Computational Performance}\label{sec:Benchmark}
Our main goal is to develop an efficient numerical method for tailoring the nonequilibrium transport of electronic systems. In order to estimate the computational effort, we perform a set of simulations for different system parameters. Numerical simulations are carried out for an initial plane wave state ($k=0$) and its time--evolved state. The quenching parameters are $W_{f}=10$ for the standard Anderson model. It is important to note that the numerical results converge to the analytical results in the thermodynamic limit for sufficiently large system sizes with $M = 1024$ Chebyshev moments. Therefore, we keep the Chebyshev moments fixed and estimate the computational cost (execution time per simulation step) and memory consumption of the numerical technique.

The main concern in numerical implementations is the scaling of the computational cost and memory requirements with respect to the size of the Hamiltonian. The computational efforts to calculate the Loschmidt echo of the quenched Anderson model with the size of the system are shown in Fig.~\ref{fig:Complexity}. We emphasize on the execution time (sec) and memory usage (GB) of the EDM (upper panels) and KPM (lower panels) techniques of C++ programming for the quench dynamics under the Anderson model. The execution time and memory usage of the EDM are approximately scaled as $N^{1.895}$ and $N^{0.98}$ for a sparse Hamiltonian matrix obtained by fitting the data as shown by the magenta dashed line in Fig.~\ref{fig:Complexity}(a) and Fig.~\ref{fig:Complexity}(b), respectively. In comparison, the polynomial-expansion scheme proves to be an efficient numerical method with a computational cost that scales linearly with the system size. More precisely, both the execution time and the memory usage scale as $ N^{1.05}$ and $N^{0.96}$, as shown by the magenta dashed line in Fig.~\ref{fig:Complexity}(c) and Fig.~\ref{fig:Complexity}(d), respectively. It can be seen that the memory consumption of the Loschmidt echo simulations by EDM and KPM scales linearly with system sizes, but the execution time caused by the diagonalization approach can be successfully circumvented by using the polynomial-expansion scheme.

The main concern of this work is to develop a linear scaling simulation technique for the investigation of the Loschmidt echo of noninteracting quantum quenched systems. However, a more fascinating road map of research is to employ the polynomial-expansion scheme for the interacting fermionic systems under quench dynamics, whose numerical effort scales exponentially.
\section{Conclusion}
We have developed an efficient numerical technique based on the  polynomial-expansion and applied it to the nonequilibrium quantum transport of noninteracting tight-binding quantum quench systems. The computational simulation of the Loschmidt echo provides new scientific insights into the nonequilibrium transport physics in the extended and localized regimes, as well as the dynamical phase transition in large-scale materials. The computational cost of applications can be greatly reduced using the polynomial-expansion approach, which scales linearly with the size of the system.

We have illustrated the applicability of the linear-scaling numerical technique to quench dynamics under quasicrystal and disordered lattices. We have verified the existence of the dynamical phase transitions under the Aubry--André model using the linear-scaling quantum transport technique, characterized by the singular behavior of the Loschmidt echo at critical times. Furthermore, the computational method has been implemented for the uncorrelated and correlated Anderson models to study many emerging and complex quantum transport phenomena that are difficult to fully address with the exact diagonalization approach. Moreover, the role of the wave vector on the quench dynamics has also been explored. We observed wave vector-independent DQPTs persistently for models that exhibit a quantum phase transition without mobility edges.
%
\section*{CRediT authorship contribution statement}
\textbf{Niaz Ali Khan:} Conceptualization, Methodology, Investigation, Original draft preparation. \textbf{Wen Chen:} Software, Data curation, Validation, Reviewing, and Editing. \textbf{Munsif Jan:} Software, Validation, Visualization, Reviewing and editing. \textbf{Gao Xianlong:} Reviewing and editing, Project administration, Supervision, Funding acquisition.
%
\section*{Declaration of competing interest}
The authors declare that they have no known competing financial interests or personal relationships that could have appeared to influence the work reported in this paper.
%
\section*{Data availability}
Data will be made available on request.
%
\begin{figure}
	\begin{centering}
		\includegraphics[scale=0.36]{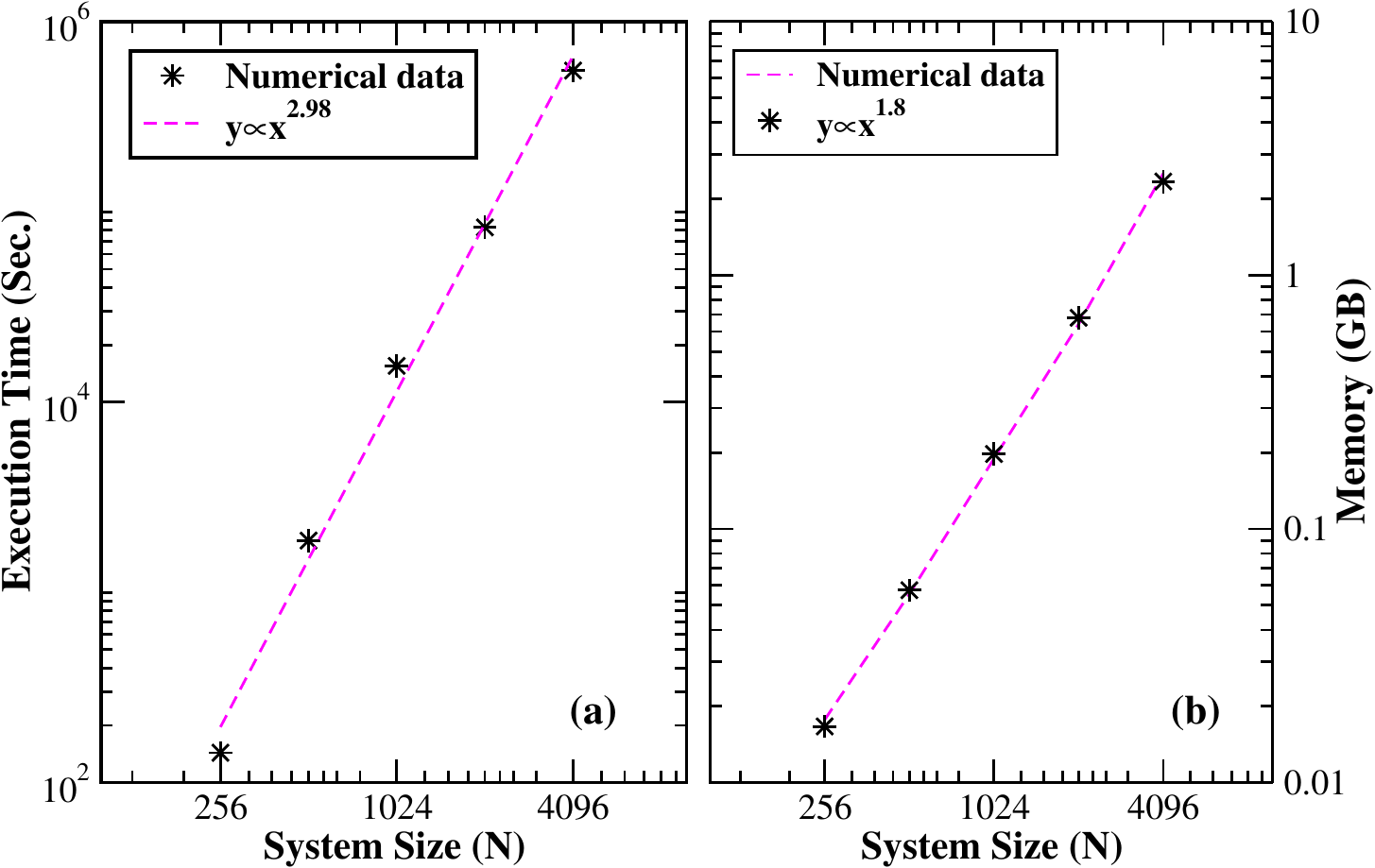}
		\par\end{centering}
	\caption{(color online) Scaling of the (a) execution time (sec) and (b) memory usage (GB) for calculating the Loschmidt echo of the Anderson model based on the exact diagonalization. Numerical computations are performed for an initial plane wave ($k=0$) quenched into a time-evolved state with $W_{f}=10$ for a single realization of disorder. The magenta dashed curves are the corresponding fits to the numerical data.\label{fig:ComplexityEDM}}
\end{figure}
\begin{acknowledgments}
N.A.K. and M.J. acknowledge the postdoctoral fellowship supported by Zhejiang Normal University under Grants No. ZC304022980 and No. ZC304022918, respectively. G.X. acknowledges support from the NSFC under Grants No. 12174346.
\end{acknowledgments}
%
\section*{Appendix}
\appendix
\section{Numerical complexity for a dense Hamiltonian matrix}\label{appendix:appendixA}
It is a well-established fact that the computational complexity of the Loschmidt echo based on the exact diagonalization method is $\mathcal{O}(N^{3})$ for a dense Hamiltonian matrix, where $N$ is the system size. In order to verify the computational cost of the method, we investigate computational efforts by calculating the Loschmidt echo for a dense Hamiltonian matrix under the quenched Anderson model, which is displayed in Fig.~\ref{fig:ComplexityEDM}. We emphasis on the execution time (sec) and memory usage (GB) of the quench dynamics under the Anderson model. The execution time and memory usage of the EDM approximately scale as $N^{2.98}$ and $N^{1.88}$ for the dense Hamiltonian matrix obtained by fitting the data as shown by the magenta dashed line in Fig.~\ref{fig:ComplexityEDM}(a) and Fig.~\ref{fig:ComplexityEDM}(b), respectively.
\bibliographystyle{apsrev4-2.bst}
\bibliography{KPMLE}

\begin{thebibliography}{55}%
\makeatletter
\providecommand \@ifxundefined [1]{%
 \@ifx{#1\undefined}
}%
\providecommand \@ifnum [1]{%
 \ifnum #1\expandafter \@firstoftwo
 \else \expandafter \@secondoftwo
 \fi
}%
\providecommand \@ifx [1]{%
 \ifx #1\expandafter \@firstoftwo
 \else \expandafter \@secondoftwo
 \fi
}%
\providecommand \natexlab [1]{#1}%
\providecommand \enquote  [1]{``#1''}%
\providecommand \bibnamefont  [1]{#1}%
\providecommand \bibfnamefont [1]{#1}%
\providecommand \citenamefont [1]{#1}%
\providecommand \href@noop [0]{\@secondoftwo}%
\providecommand \href [0]{\begingroup \@sanitize@url \@href}%
\providecommand \@href[1]{\@@startlink{#1}\@@href}%
\providecommand \@@href[1]{\endgroup#1\@@endlink}%
\providecommand \@sanitize@url [0]{\catcode `\\12\catcode `\$12\catcode
  `\&12\catcode `\#12\catcode `\^12\catcode `\_12\catcode `\%12\relax}%
\providecommand \@@startlink[1]{}%
\providecommand \@@endlink[0]{}%
\providecommand \url  [0]{\begingroup\@sanitize@url \@url }%
\providecommand \@url [1]{\endgroup\@href {#1}{\urlprefix }}%
\providecommand \urlprefix  [0]{URL }%
\providecommand \Eprint [0]{\href }%
\providecommand \doibase [0]{https://doi.org/}%
\providecommand \selectlanguage [0]{\@gobble}%
\providecommand \bibinfo  [0]{\@secondoftwo}%
\providecommand \bibfield  [0]{\@secondoftwo}%
\providecommand \translation [1]{[#1]}%
\providecommand \BibitemOpen [0]{}%
\providecommand \bibitemStop [0]{}%
\providecommand \bibitemNoStop [0]{.\EOS\space}%
\providecommand \EOS [0]{\spacefactor3000\relax}%
\providecommand \BibitemShut  [1]{\csname bibitem#1\endcsname}%
\let\auto@bib@innerbib\@empty
\bibitem [{\citenamefont {Wilkinson}(1988)}]{Wilkinson1988}%
  \BibitemOpen
  \bibfield  {author} {\bibinfo {author} {\bibfnamefont {J.~H.}\ \bibnamefont
  {Wilkinson}},\ }\href {https://doi.org/doi:10.1017/S0013091500012104} {\emph
  {\bibinfo {title} {The Algebraic Eigenvalue Problem}}}\ (\bibinfo
  {publisher} {Oxford University Press, Inc., USA},\ \bibinfo {year}
  {1988})\BibitemShut {NoStop}%
\bibitem [{\citenamefont {Yin}\ \emph {et~al.}(2018)\citenamefont {Yin},
  \citenamefont {Chen}, \citenamefont {Xianlong},\ and\ \citenamefont
  {Wang}}]{Yin2018}%
  \BibitemOpen
  \bibfield  {author} {\bibinfo {author} {\bibfnamefont {H.}~\bibnamefont
  {Yin}}, \bibinfo {author} {\bibfnamefont {S.}~\bibnamefont {Chen}}, \bibinfo
  {author} {\bibfnamefont {G.}~\bibnamefont {Xianlong}}, \ and\ \bibinfo
  {author} {\bibfnamefont {P.}~\bibnamefont {Wang}},\ }\href
  {https://doi.org/10.1103/PhysRevA.97.033624} {\bibfield  {journal} {\bibinfo
  {journal} {Phys. Rev. A}\ }\textbf {\bibinfo {volume} {97}},\ \bibinfo
  {pages} {033624} (\bibinfo {year} {2018})}\BibitemShut {NoStop}%
\bibitem [{\citenamefont {Fadel}\ \emph {et~al.}(2021)\citenamefont {Fadel},
  \citenamefont {Usui}, \citenamefont {Huber}, \citenamefont {Friis},\ and\
  \citenamefont {Vitagliano}}]{Fadel2021}%
  \BibitemOpen
  \bibfield  {author} {\bibinfo {author} {\bibfnamefont {M.}~\bibnamefont
  {Fadel}}, \bibinfo {author} {\bibfnamefont {A.}~\bibnamefont {Usui}},
  \bibinfo {author} {\bibfnamefont {M.}~\bibnamefont {Huber}}, \bibinfo
  {author} {\bibfnamefont {N.}~\bibnamefont {Friis}}, \ and\ \bibinfo {author}
  {\bibfnamefont {G.}~\bibnamefont {Vitagliano}},\ }\href
  {https://doi.org/10.1103/PhysRevLett.127.010401} {\bibfield  {journal}
  {\bibinfo  {journal} {Phys. Rev. Lett.}\ }\textbf {\bibinfo {volume} {127}},\
  \bibinfo {pages} {010401} (\bibinfo {year} {2021})}\BibitemShut {NoStop}%
\bibitem [{\citenamefont {Wei\ss{}e}\ \emph {et~al.}(2006)\citenamefont
  {Wei\ss{}e}, \citenamefont {Wellein}, \citenamefont {Alvermann},\ and\
  \citenamefont {Fehske}}]{Weibe2006}%
  \BibitemOpen
  \bibfield  {author} {\bibinfo {author} {\bibfnamefont {A.}~\bibnamefont
  {Wei\ss{}e}}, \bibinfo {author} {\bibfnamefont {G.}~\bibnamefont {Wellein}},
  \bibinfo {author} {\bibfnamefont {A.}~\bibnamefont {Alvermann}}, \ and\
  \bibinfo {author} {\bibfnamefont {H.}~\bibnamefont {Fehske}},\ }\href
  {https://doi.org/10.1103/RevModPhys.78.275} {\bibfield  {journal} {\bibinfo
  {journal} {Rev. Mod. Phys.}\ }\textbf {\bibinfo {volume} {78}},\ \bibinfo
  {pages} {275} (\bibinfo {year} {2006})}\BibitemShut {NoStop}%
\bibitem [{\citenamefont {João}\ \emph {et~al.}(2020)\citenamefont {João},
  \citenamefont {Andelkovic}, \citenamefont {Covaci}, \citenamefont
  {Rappoport}, \citenamefont {Lopes},\ and\ \citenamefont
  {Ferreira}}]{Simao2020}%
  \BibitemOpen
  \bibfield  {author} {\bibinfo {author} {\bibfnamefont {S.~M.}\ \bibnamefont
  {João}}, \bibinfo {author} {\bibfnamefont {M.}~\bibnamefont {Andelkovic}},
  \bibinfo {author} {\bibfnamefont {L.}~\bibnamefont {Covaci}}, \bibinfo
  {author} {\bibfnamefont {T.~G.}\ \bibnamefont {Rappoport}}, \bibinfo {author}
  {\bibfnamefont {J.~M. V.~P.}\ \bibnamefont {Lopes}}, \ and\ \bibinfo {author}
  {\bibfnamefont {A.}~\bibnamefont {Ferreira}},\ }\href
  {https://doi.org/10.1098/rsos.191809} {\bibfield  {journal} {\bibinfo
  {journal} {R. Soc. Open Sci.}\ }\textbf {\bibinfo {volume} {7}},\ \bibinfo
  {pages} {191809} (\bibinfo {year} {2020})}\BibitemShut {NoStop}%
\bibitem [{\citenamefont {Khan}\ and\ \citenamefont {Amin}(2021)}]{Niaz2021}%
  \BibitemOpen
  \bibfield  {author} {\bibinfo {author} {\bibfnamefont {N.~A.}\ \bibnamefont
  {Khan}}\ and\ \bibinfo {author} {\bibfnamefont {S.~T.}\ \bibnamefont
  {Amin}},\ }\href {https://doi.org/10.1088/1402-4896/abe322} {\bibfield
  {journal} {\bibinfo  {journal} {Phys. Scr.}\ }\textbf {\bibinfo {volume}
  {96}},\ \bibinfo {pages} {045812} (\bibinfo {year} {2021})}\BibitemShut
  {NoStop}%
\bibitem [{\citenamefont {Fan}\ \emph {et~al.}(2021)\citenamefont {Fan},
  \citenamefont {Garcia}, \citenamefont {Cummings}, \citenamefont
  {Barrios-Vargas}, \citenamefont {Panhans}, \citenamefont {Harju},
  \citenamefont {Ortmann},\ and\ \citenamefont {Roche}}]{FAN2021}%
  \BibitemOpen
  \bibfield  {author} {\bibinfo {author} {\bibfnamefont {Z.}~\bibnamefont
  {Fan}}, \bibinfo {author} {\bibfnamefont {J.~H.}\ \bibnamefont {Garcia}},
  \bibinfo {author} {\bibfnamefont {A.~W.}\ \bibnamefont {Cummings}}, \bibinfo
  {author} {\bibfnamefont {J.~E.}\ \bibnamefont {Barrios-Vargas}}, \bibinfo
  {author} {\bibfnamefont {M.}~\bibnamefont {Panhans}}, \bibinfo {author}
  {\bibfnamefont {A.}~\bibnamefont {Harju}}, \bibinfo {author} {\bibfnamefont
  {F.}~\bibnamefont {Ortmann}}, \ and\ \bibinfo {author} {\bibfnamefont
  {S.}~\bibnamefont {Roche}},\ }\href
  {https://doi.org/https://doi.org/10.1016/j.physrep.2020.12.001} {\bibfield
  {journal} {\bibinfo  {journal} {Phys. Rep.}\ }\textbf {\bibinfo {volume}
  {903}},\ \bibinfo {pages} {1} (\bibinfo {year} {2021})}\BibitemShut {NoStop}%
\bibitem [{\citenamefont {João}\ \emph {et~al.}(2022)\citenamefont {João},
  \citenamefont {Lopes},\ and\ \citenamefont {Ferreira}}]{Simao2022}%
  \BibitemOpen
  \bibfield  {author} {\bibinfo {author} {\bibfnamefont {S.~M.}\ \bibnamefont
  {João}}, \bibinfo {author} {\bibfnamefont {J.~M. V.~P.}\ \bibnamefont
  {Lopes}}, \ and\ \bibinfo {author} {\bibfnamefont {A.}~\bibnamefont
  {Ferreira}},\ }\href {https://doi.org/10.1088/2515-7639/ac91f9} {\bibfield
  {journal} {\bibinfo  {journal} {Journal of Physics: Materials}\ }\textbf
  {\bibinfo {volume} {5}},\ \bibinfo {pages} {045002} (\bibinfo {year}
  {2022})}\BibitemShut {NoStop}%
\bibitem [{\citenamefont {Bjelcic}\ \emph {et~al.}(2022)\citenamefont
  {Bjelcic}, \citenamefont {Niksic},\ and\ \citenamefont
  {Drmac}}]{BJELCIC2022108477}%
  \BibitemOpen
  \bibfield  {author} {\bibinfo {author} {\bibfnamefont {A.}~\bibnamefont
  {Bjelcic}}, \bibinfo {author} {\bibfnamefont {T.}~\bibnamefont {Niksic}}, \
  and\ \bibinfo {author} {\bibfnamefont {Z.}~\bibnamefont {Drmac}},\ }\href
  {https://doi.org/10.1016/j.cpc.2022.108477} {\bibfield  {journal} {\bibinfo
  {journal} {Comput. Phys. Commun.}\ }\textbf {\bibinfo {volume} {280}},\
  \bibinfo {pages} {108477} (\bibinfo {year} {2022})}\BibitemShut {NoStop}%
\bibitem [{\citenamefont {Sobczyk}\ and\ \citenamefont
  {Roggero}(2022)}]{Sobczyk2022}%
  \BibitemOpen
  \bibfield  {author} {\bibinfo {author} {\bibfnamefont {J.~E.}\ \bibnamefont
  {Sobczyk}}\ and\ \bibinfo {author} {\bibfnamefont {A.}~\bibnamefont
  {Roggero}},\ }\href {https://doi.org/10.1103/PhysRevE.105.055310} {\bibfield
  {journal} {\bibinfo  {journal} {Phys. Rev. E}\ }\textbf {\bibinfo {volume}
  {105}},\ \bibinfo {pages} {055310} (\bibinfo {year} {2022})}\BibitemShut
  {NoStop}%
\bibitem [{\citenamefont {de~Castro}\ \emph {et~al.}(2023)\citenamefont
  {de~Castro}, \citenamefont {Ferreira},\ and\ \citenamefont
  {Bahamon}}]{Castro2023}%
  \BibitemOpen
  \bibfield  {author} {\bibinfo {author} {\bibfnamefont {S.~G.}\ \bibnamefont
  {de~Castro}}, \bibinfo {author} {\bibfnamefont {A.}~\bibnamefont {Ferreira}},
  \ and\ \bibinfo {author} {\bibfnamefont {D.~A.}\ \bibnamefont {Bahamon}},\
  }\href {https://doi.org/10.1103/PhysRevB.107.045418} {\bibfield  {journal}
  {\bibinfo  {journal} {Phys. Rev. B}\ }\textbf {\bibinfo {volume} {107}},\
  \bibinfo {pages} {045418} (\bibinfo {year} {2023})}\BibitemShut {NoStop}%
\bibitem [{\citenamefont {Li}\ \emph {et~al.}(2023)\citenamefont {Li},
  \citenamefont {Zhan}, \citenamefont {Kuang}, \citenamefont {Li},\ and\
  \citenamefont {Yuan}}]{Li2023}%
  \BibitemOpen
  \bibfield  {author} {\bibinfo {author} {\bibfnamefont {Y.}~\bibnamefont
  {Li}}, \bibinfo {author} {\bibfnamefont {Z.}~\bibnamefont {Zhan}}, \bibinfo
  {author} {\bibfnamefont {X.}~\bibnamefont {Kuang}}, \bibinfo {author}
  {\bibfnamefont {Y.}~\bibnamefont {Li}}, \ and\ \bibinfo {author}
  {\bibfnamefont {S.}~\bibnamefont {Yuan}},\ }\href
  {https://doi.org/https://doi.org/10.1016/j.cpc.2022.108632} {\bibfield
  {journal} {\bibinfo  {journal} {Comput. Phys. Commun.}\ }\textbf {\bibinfo
  {volume} {285}},\ \bibinfo {pages} {108632} (\bibinfo {year}
  {2023})}\BibitemShut {NoStop}%
\bibitem [{\citenamefont {Zhao}\ \emph {et~al.}(2023)\citenamefont {Zhao},
  \citenamefont {Ding},\ and\ \citenamefont {Yang}}]{Zhao2023}%
  \BibitemOpen
  \bibfield  {author} {\bibinfo {author} {\bibfnamefont {P.-Y.}\ \bibnamefont
  {Zhao}}, \bibinfo {author} {\bibfnamefont {K.}~\bibnamefont {Ding}}, \ and\
  \bibinfo {author} {\bibfnamefont {S.}~\bibnamefont {Yang}},\ }\href
  {https://doi.org/10.1103/PhysRevResearch.5.023026} {\bibfield  {journal}
  {\bibinfo  {journal} {Phys. Rev. Res.}\ }\textbf {\bibinfo {volume} {5}},\
  \bibinfo {pages} {023026} (\bibinfo {year} {2023})}\BibitemShut {NoStop}%
\bibitem [{\citenamefont {Elstner}\ \emph {et~al.}(1998)\citenamefont
  {Elstner}, \citenamefont {Porezag}, \citenamefont {Jungnickel}, \citenamefont
  {Elsner}, \citenamefont {Haugk}, \citenamefont {Frauenheim}, \citenamefont
  {Suhai},\ and\ \citenamefont {Seifert}}]{Elstner1998}%
  \BibitemOpen
  \bibfield  {author} {\bibinfo {author} {\bibfnamefont {M.}~\bibnamefont
  {Elstner}}, \bibinfo {author} {\bibfnamefont {D.}~\bibnamefont {Porezag}},
  \bibinfo {author} {\bibfnamefont {G.}~\bibnamefont {Jungnickel}}, \bibinfo
  {author} {\bibfnamefont {J.}~\bibnamefont {Elsner}}, \bibinfo {author}
  {\bibfnamefont {M.}~\bibnamefont {Haugk}}, \bibinfo {author} {\bibfnamefont
  {T.}~\bibnamefont {Frauenheim}}, \bibinfo {author} {\bibfnamefont
  {S.}~\bibnamefont {Suhai}}, \ and\ \bibinfo {author} {\bibfnamefont
  {G.}~\bibnamefont {Seifert}},\ }\href
  {https://doi.org/10.1103/PhysRevB.58.7260} {\bibfield  {journal} {\bibinfo
  {journal} {Phys. Rev. B}\ }\textbf {\bibinfo {volume} {58}},\ \bibinfo
  {pages} {7260} (\bibinfo {year} {1998})}\BibitemShut {NoStop}%
\bibitem [{\citenamefont {Zen}\ \emph {et~al.}(2018)\citenamefont {Zen},
  \citenamefont {Brandenburg}, \citenamefont {Klimeš}, \citenamefont
  {Tkatchenko}, \citenamefont {Alfè},\ and\ \citenamefont
  {Michaelides}}]{Zen2018}%
  \BibitemOpen
  \bibfield  {author} {\bibinfo {author} {\bibfnamefont {A.}~\bibnamefont
  {Zen}}, \bibinfo {author} {\bibfnamefont {J.~G.}\ \bibnamefont
  {Brandenburg}}, \bibinfo {author} {\bibfnamefont {J.}~\bibnamefont
  {Klimeš}}, \bibinfo {author} {\bibfnamefont {A.}~\bibnamefont {Tkatchenko}},
  \bibinfo {author} {\bibfnamefont {D.}~\bibnamefont {Alfè}}, \ and\ \bibinfo
  {author} {\bibfnamefont {A.}~\bibnamefont {Michaelides}},\ }\href
  {https://doi.org/10.1073/pnas.1715434115} {\bibfield  {journal} {\bibinfo
  {journal} {Proc. Natl Acad. Sci.}\ }\textbf {\bibinfo {volume} {115}},\
  \bibinfo {pages} {1724} (\bibinfo {year} {2018})}\BibitemShut {NoStop}%
\bibitem [{\citenamefont {Schollw\"ock}(2005)}]{Schollwock2005}%
  \BibitemOpen
  \bibfield  {author} {\bibinfo {author} {\bibfnamefont {U.}~\bibnamefont
  {Schollw\"ock}},\ }\href {https://doi.org/10.1103/RevModPhys.77.259}
  {\bibfield  {journal} {\bibinfo  {journal} {Rev. Mod. Phys.}\ }\textbf
  {\bibinfo {volume} {77}},\ \bibinfo {pages} {259} (\bibinfo {year}
  {2005})}\BibitemShut {NoStop}%
\bibitem [{\citenamefont {Dreizler}\ and\ \citenamefont
  {Gross}(1990)}]{Dreizler1990}%
  \BibitemOpen
  \bibfield  {author} {\bibinfo {author} {\bibfnamefont {R.~M.}\ \bibnamefont
  {Dreizler}}\ and\ \bibinfo {author} {\bibfnamefont {E.~K.~U.}\ \bibnamefont
  {Gross}},\ }\href {https://doi.org/doi.org/10.1007/978-3-642-86105-5} {\emph
  {\bibinfo {title} {Density Functional Theory: An Approach to the Quantum
  Many-Body Problem}}}\ (\bibinfo  {publisher} {Springer Berlin, Heidelberg},\
  \bibinfo {year} {1990})\BibitemShut {NoStop}%
\bibitem [{\citenamefont {Onida}\ \emph {et~al.}(2002)\citenamefont {Onida},
  \citenamefont {Reining},\ and\ \citenamefont {Rubio}}]{Onida2002}%
  \BibitemOpen
  \bibfield  {author} {\bibinfo {author} {\bibfnamefont {G.}~\bibnamefont
  {Onida}}, \bibinfo {author} {\bibfnamefont {L.}~\bibnamefont {Reining}}, \
  and\ \bibinfo {author} {\bibfnamefont {A.}~\bibnamefont {Rubio}},\ }\href
  {https://doi.org/10.1103/RevModPhys.74.601} {\bibfield  {journal} {\bibinfo
  {journal} {Rev. Mod. Phys.}\ }\textbf {\bibinfo {volume} {74}},\ \bibinfo
  {pages} {601} (\bibinfo {year} {2002})}\BibitemShut {NoStop}%
\bibitem [{\citenamefont {Heyl}(2015)}]{Heyl2015}%
  \BibitemOpen
  \bibfield  {author} {\bibinfo {author} {\bibfnamefont {M.}~\bibnamefont
  {Heyl}},\ }\href {https://link.aps.org/doi/10.1103/PhysRevLett.115.140602}
  {\bibfield  {journal} {\bibinfo  {journal} {Phys. Rev. Lett.}\ }\textbf
  {\bibinfo {volume} {115}},\ \bibinfo {pages} {140602} (\bibinfo {year}
  {2015})}\BibitemShut {NoStop}%
\bibitem [{\citenamefont {Heyl}(2018)}]{Heyl2018}%
  \BibitemOpen
  \bibfield  {author} {\bibinfo {author} {\bibfnamefont {M.}~\bibnamefont
  {Heyl}},\ }\href {https://doi.org/10.1088/1361-6633/aaaf9a} {\bibfield
  {journal} {\bibinfo  {journal} {Rep. Prog. Phys.}\ }\textbf {\bibinfo
  {volume} {81}},\ \bibinfo {pages} {054001} (\bibinfo {year}
  {2018})}\BibitemShut {NoStop}%
\bibitem [{\citenamefont {Sadrzadeh}\ \emph {et~al.}(2021)\citenamefont
  {Sadrzadeh}, \citenamefont {Jafari},\ and\ \citenamefont
  {Langari}}]{Sadrzadeh2021}%
  \BibitemOpen
  \bibfield  {author} {\bibinfo {author} {\bibfnamefont {M.}~\bibnamefont
  {Sadrzadeh}}, \bibinfo {author} {\bibfnamefont {R.}~\bibnamefont {Jafari}}, \
  and\ \bibinfo {author} {\bibfnamefont {A.}~\bibnamefont {Langari}},\ }\href
  {https://doi.org/10.1103/PhysRevB.103.144305} {\bibfield  {journal} {\bibinfo
   {journal} {Phys. Rev. B}\ }\textbf {\bibinfo {volume} {103}},\ \bibinfo
  {pages} {144305} (\bibinfo {year} {2021})}\BibitemShut {NoStop}%
\bibitem [{\citenamefont {Jafari}\ \emph {et~al.}(2022)\citenamefont {Jafari},
  \citenamefont {Akbari}, \citenamefont {Mishra},\ and\ \citenamefont
  {Johannesson}}]{Jafari2022}%
  \BibitemOpen
  \bibfield  {author} {\bibinfo {author} {\bibfnamefont {R.}~\bibnamefont
  {Jafari}}, \bibinfo {author} {\bibfnamefont {A.}~\bibnamefont {Akbari}},
  \bibinfo {author} {\bibfnamefont {U.}~\bibnamefont {Mishra}}, \ and\ \bibinfo
  {author} {\bibfnamefont {H.}~\bibnamefont {Johannesson}},\ }\href
  {https://doi.org/10.1103/PhysRevB.105.094311} {\bibfield  {journal} {\bibinfo
   {journal} {Phys. Rev. B}\ }\textbf {\bibinfo {volume} {105}},\ \bibinfo
  {pages} {094311} (\bibinfo {year} {2022})}\BibitemShut {NoStop}%
\bibitem [{\citenamefont {De~Nicola}\ \emph {et~al.}(2022)\citenamefont
  {De~Nicola}, \citenamefont {Michailidis},\ and\ \citenamefont
  {Serbyn}}]{Nicola2022}%
  \BibitemOpen
  \bibfield  {author} {\bibinfo {author} {\bibfnamefont {S.}~\bibnamefont
  {De~Nicola}}, \bibinfo {author} {\bibfnamefont {A.~A.}\ \bibnamefont
  {Michailidis}}, \ and\ \bibinfo {author} {\bibfnamefont {M.}~\bibnamefont
  {Serbyn}},\ }\href {https://doi.org/10.1103/PhysRevB.105.165149} {\bibfield
  {journal} {\bibinfo  {journal} {Phys. Rev. B}\ }\textbf {\bibinfo {volume}
  {105}},\ \bibinfo {pages} {165149} (\bibinfo {year} {2022})}\BibitemShut
  {NoStop}%
\bibitem [{\citenamefont {Hamazaki}(2021)}]{Hamazaki2021}%
  \BibitemOpen
  \bibfield  {author} {\bibinfo {author} {\bibfnamefont {R.}~\bibnamefont
  {Hamazaki}},\ }\href {https://doi.org/10.1038/s41467-021-25355-3} {\bibfield
  {journal} {\bibinfo  {journal} {Nat. Commun.}\ }\textbf {\bibinfo {volume}
  {12}},\ \bibinfo {pages} {5108} (\bibinfo {year} {2021})}\BibitemShut
  {NoStop}%
\bibitem [{\citenamefont {Van~Damme}\ \emph {et~al.}(2022)\citenamefont
  {Van~Damme}, \citenamefont {Zache}, \citenamefont {Banerjee}, \citenamefont
  {Hauke},\ and\ \citenamefont {Halimeh}}]{Damme2022}%
  \BibitemOpen
  \bibfield  {author} {\bibinfo {author} {\bibfnamefont {M.}~\bibnamefont
  {Van~Damme}}, \bibinfo {author} {\bibfnamefont {T.~V.}\ \bibnamefont
  {Zache}}, \bibinfo {author} {\bibfnamefont {D.}~\bibnamefont {Banerjee}},
  \bibinfo {author} {\bibfnamefont {P.}~\bibnamefont {Hauke}}, \ and\ \bibinfo
  {author} {\bibfnamefont {J.~C.}\ \bibnamefont {Halimeh}},\ }\href
  {https://doi.org/10.1103/PhysRevB.106.245110} {\bibfield  {journal} {\bibinfo
   {journal} {Phys. Rev. B}\ }\textbf {\bibinfo {volume} {106}},\ \bibinfo
  {pages} {245110} (\bibinfo {year} {2022})}\BibitemShut {NoStop}%
\bibitem [{\citenamefont {Peotta}\ \emph {et~al.}(2021)\citenamefont {Peotta},
  \citenamefont {Brange}, \citenamefont {Deger}, \citenamefont {Ojanen},\ and\
  \citenamefont {Flindt}}]{Peotta2021}%
  \BibitemOpen
  \bibfield  {author} {\bibinfo {author} {\bibfnamefont {S.}~\bibnamefont
  {Peotta}}, \bibinfo {author} {\bibfnamefont {F.}~\bibnamefont {Brange}},
  \bibinfo {author} {\bibfnamefont {A.}~\bibnamefont {Deger}}, \bibinfo
  {author} {\bibfnamefont {T.}~\bibnamefont {Ojanen}}, \ and\ \bibinfo {author}
  {\bibfnamefont {C.}~\bibnamefont {Flindt}},\ }\href
  {https://doi.org/10.1103/PhysRevX.11.041018} {\bibfield  {journal} {\bibinfo
  {journal} {Phys. Rev. X}\ }\textbf {\bibinfo {volume} {11}},\ \bibinfo
  {pages} {041018} (\bibinfo {year} {2021})}\BibitemShut {NoStop}%
\bibitem [{\citenamefont {Zurek}\ \emph {et~al.}(2005)\citenamefont {Zurek},
  \citenamefont {Dorner},\ and\ \citenamefont {Zoller}}]{Zurek2005}%
  \BibitemOpen
  \bibfield  {author} {\bibinfo {author} {\bibfnamefont {W.~H.}\ \bibnamefont
  {Zurek}}, \bibinfo {author} {\bibfnamefont {U.}~\bibnamefont {Dorner}}, \
  and\ \bibinfo {author} {\bibfnamefont {P.}~\bibnamefont {Zoller}},\ }\href
  {https://doi.org/10.1103/PhysRevLett.95.105701} {\bibfield  {journal}
  {\bibinfo  {journal} {Phys. Rev. Lett.}\ }\textbf {\bibinfo {volume} {95}},\
  \bibinfo {pages} {105701} (\bibinfo {year} {2005})}\BibitemShut {NoStop}%
\bibitem [{\citenamefont {Chen}\ \emph {et~al.}(2011)\citenamefont {Chen},
  \citenamefont {White}, \citenamefont {Borries},\ and\ \citenamefont
  {DeMarco}}]{Chen2011}%
  \BibitemOpen
  \bibfield  {author} {\bibinfo {author} {\bibfnamefont {D.}~\bibnamefont
  {Chen}}, \bibinfo {author} {\bibfnamefont {M.}~\bibnamefont {White}},
  \bibinfo {author} {\bibfnamefont {C.}~\bibnamefont {Borries}}, \ and\
  \bibinfo {author} {\bibfnamefont {B.}~\bibnamefont {DeMarco}},\ }\href
  {https://doi.org/10.1103/PhysRevLett.106.235304} {\bibfield  {journal}
  {\bibinfo  {journal} {Phys. Rev. Lett.}\ }\textbf {\bibinfo {volume} {106}},\
  \bibinfo {pages} {235304} (\bibinfo {year} {2011})}\BibitemShut {NoStop}%
\bibitem [{\citenamefont {Yang}\ \emph {et~al.}(2017)\citenamefont {Yang},
  \citenamefont {Wang}, \citenamefont {Wang}, \citenamefont {Xianlong},\ and\
  \citenamefont {Chen}}]{Yang2017}%
  \BibitemOpen
  \bibfield  {author} {\bibinfo {author} {\bibfnamefont {C.}~\bibnamefont
  {Yang}}, \bibinfo {author} {\bibfnamefont {Y.}~\bibnamefont {Wang}}, \bibinfo
  {author} {\bibfnamefont {P.}~\bibnamefont {Wang}}, \bibinfo {author}
  {\bibfnamefont {G.}~\bibnamefont {Xianlong}}, \ and\ \bibinfo {author}
  {\bibfnamefont {S.}~\bibnamefont {Chen}},\ }\href
  {https://link.aps.org/doi/10.1103/PhysRevB.95.184201} {\bibfield  {journal}
  {\bibinfo  {journal} {Phys. Rev. B}\ }\textbf {\bibinfo {volume} {95}},\
  \bibinfo {pages} {184201} (\bibinfo {year} {2017})}\BibitemShut {NoStop}%
\bibitem [{\citenamefont {Xu}\ and\ \citenamefont {Chen}(2021)}]{Xu2021}%
  \BibitemOpen
  \bibfield  {author} {\bibinfo {author} {\bibfnamefont {Z.}~\bibnamefont
  {Xu}}\ and\ \bibinfo {author} {\bibfnamefont {S.}~\bibnamefont {Chen}},\
  }\href {https://doi.org/10.1103/PhysRevA.103.043325} {\bibfield  {journal}
  {\bibinfo  {journal} {Phys. Rev. A}\ }\textbf {\bibinfo {volume} {103}},\
  \bibinfo {pages} {043325} (\bibinfo {year} {2021})}\BibitemShut {NoStop}%
\bibitem [{\citenamefont {Xu}\ \emph {et~al.}(2020)\citenamefont {Xu},
  \citenamefont {Sun}, \citenamefont {Liu}, \citenamefont {Zhang},
  \citenamefont {Li}, \citenamefont {Dong}, \citenamefont {Ren}, \citenamefont
  {Zhang}, \citenamefont {Nori}, \citenamefont {Zheng}, \citenamefont {Fan},\
  and\ \citenamefont {Wang}}]{Xu2020}%
  \BibitemOpen
  \bibfield  {author} {\bibinfo {author} {\bibfnamefont {K.}~\bibnamefont
  {Xu}}, \bibinfo {author} {\bibfnamefont {Z.-H.}\ \bibnamefont {Sun}},
  \bibinfo {author} {\bibfnamefont {W.}~\bibnamefont {Liu}}, \bibinfo {author}
  {\bibfnamefont {Y.-R.}\ \bibnamefont {Zhang}}, \bibinfo {author}
  {\bibfnamefont {H.}~\bibnamefont {Li}}, \bibinfo {author} {\bibfnamefont
  {H.}~\bibnamefont {Dong}}, \bibinfo {author} {\bibfnamefont {W.}~\bibnamefont
  {Ren}}, \bibinfo {author} {\bibfnamefont {P.}~\bibnamefont {Zhang}}, \bibinfo
  {author} {\bibfnamefont {F.}~\bibnamefont {Nori}}, \bibinfo {author}
  {\bibfnamefont {D.}~\bibnamefont {Zheng}}, \bibinfo {author} {\bibfnamefont
  {H.}~\bibnamefont {Fan}}, \ and\ \bibinfo {author} {\bibfnamefont
  {H.}~\bibnamefont {Wang}},\ }\href
  {https://www.science.org/doi/10.1126/sciadv.aba4935} {\bibfield  {journal}
  {\bibinfo  {journal} {Sci. Adv.}\ }\textbf {\bibinfo {volume} {6}},\ \bibinfo
  {pages} {eaba4935} (\bibinfo {year} {2020})}\BibitemShut {NoStop}%
\bibitem [{\citenamefont {Tong}\ \emph {et~al.}(2021)\citenamefont {Tong},
  \citenamefont {Meng}, \citenamefont {Jiang}, \citenamefont {Lee},
  \citenamefont {Neto},\ and\ \citenamefont {Xianlong}}]{Tong2021}%
  \BibitemOpen
  \bibfield  {author} {\bibinfo {author} {\bibfnamefont {X.}~\bibnamefont
  {Tong}}, \bibinfo {author} {\bibfnamefont {Y.-M.}\ \bibnamefont {Meng}},
  \bibinfo {author} {\bibfnamefont {X.}~\bibnamefont {Jiang}}, \bibinfo
  {author} {\bibfnamefont {C.}~\bibnamefont {Lee}}, \bibinfo {author}
  {\bibfnamefont {G.~D. d.~M.}\ \bibnamefont {Neto}}, \ and\ \bibinfo {author}
  {\bibfnamefont {G.}~\bibnamefont {Xianlong}},\ }\href
  {https://doi.org/10.1103/PhysRevB.103.104202} {\bibfield  {journal} {\bibinfo
   {journal} {Phys. Rev. B}\ }\textbf {\bibinfo {volume} {103}},\ \bibinfo
  {pages} {104202} (\bibinfo {year} {2021})}\BibitemShut {NoStop}%
\bibitem [{\citenamefont {Khan}\ \emph
  {et~al.}(2023{\natexlab{a}})\citenamefont {Khan}, \citenamefont {Wang},
  \citenamefont {Jan},\ and\ \citenamefont {Xianlong}}]{Niaz2023cDQPT}%
  \BibitemOpen
  \bibfield  {author} {\bibinfo {author} {\bibfnamefont {N.~A.}\ \bibnamefont
  {Khan}}, \bibinfo {author} {\bibfnamefont {P.}~\bibnamefont {Wang}}, \bibinfo
  {author} {\bibfnamefont {M.}~\bibnamefont {Jan}}, \ and\ \bibinfo {author}
  {\bibfnamefont {G.}~\bibnamefont {Xianlong}},\ }\href
  {https://doi.org/10.1038/s41598-023-36564-9} {\bibfield  {journal} {\bibinfo
  {journal} {Sci. Rep.}\ }\textbf {\bibinfo {volume} {13}},\ \bibinfo {pages}
  {9470} (\bibinfo {year} {2023}{\natexlab{a}})}\BibitemShut {NoStop}%
\bibitem [{\citenamefont {Naldesi}\ \emph {et~al.}(2016)\citenamefont
  {Naldesi}, \citenamefont {Ercolessi},\ and\ \citenamefont
  {Roscilde}}]{Naldesi2016}%
  \BibitemOpen
  \bibfield  {author} {\bibinfo {author} {\bibfnamefont {P.}~\bibnamefont
  {Naldesi}}, \bibinfo {author} {\bibfnamefont {E.}~\bibnamefont {Ercolessi}},
  \ and\ \bibinfo {author} {\bibfnamefont {T.}~\bibnamefont {Roscilde}},\
  }\href {https://doi.org/10.21468/SciPostPhys.1.1.010} {\bibfield  {journal}
  {\bibinfo  {journal} {SciPost Phys.}\ }\textbf {\bibinfo {volume} {1}},\
  \bibinfo {pages} {010} (\bibinfo {year} {2016})}\BibitemShut {NoStop}%
\bibitem [{\citenamefont {Wong}\ and\ \citenamefont {Yu}(2022)}]{Wong2022}%
  \BibitemOpen
  \bibfield  {author} {\bibinfo {author} {\bibfnamefont {C.}~\bibnamefont
  {Wong}}\ and\ \bibinfo {author} {\bibfnamefont {W.~C.}\ \bibnamefont {Yu}},\
  }\href {https://doi.org/10.1103/PhysRevB.105.174307} {\bibfield  {journal}
  {\bibinfo  {journal} {Phys. Rev. B}\ }\textbf {\bibinfo {volume} {105}},\
  \bibinfo {pages} {174307} (\bibinfo {year} {2022})}\BibitemShut {NoStop}%
\bibitem [{\citenamefont {Khan}\ \emph
  {et~al.}(2023{\natexlab{b}})\citenamefont {Khan}, \citenamefont {Wei},
  \citenamefont {Cheng}, \citenamefont {Jan},\ and\ \citenamefont
  {Xianlong}}]{Niaz2023dDQPT}%
  \BibitemOpen
  \bibfield  {author} {\bibinfo {author} {\bibfnamefont {N.~A.}\ \bibnamefont
  {Khan}}, \bibinfo {author} {\bibfnamefont {X.}~\bibnamefont {Wei}}, \bibinfo
  {author} {\bibfnamefont {S.}~\bibnamefont {Cheng}}, \bibinfo {author}
  {\bibfnamefont {M.}~\bibnamefont {Jan}}, \ and\ \bibinfo {author}
  {\bibfnamefont {G.}~\bibnamefont {Xianlong}},\ }\href
  {https://doi.org/https://doi.org/10.1016/j.physleta.2023.128880} {\bibfield
  {journal} {\bibinfo  {journal} {Phys. Lett. A}\ }\textbf {\bibinfo {volume}
  {475}},\ \bibinfo {pages} {128880} (\bibinfo {year}
  {2023}{\natexlab{b}})}\BibitemShut {NoStop}%
\bibitem [{\citenamefont {Zou}\ and\ \citenamefont {Ding}(2023)}]{Zou2023}%
  \BibitemOpen
  \bibfield  {author} {\bibinfo {author} {\bibfnamefont {Y.-T.}\ \bibnamefont
  {Zou}}\ and\ \bibinfo {author} {\bibfnamefont {C.}~\bibnamefont {Ding}},\
  }\href {https://doi.org/10.1103/PhysRevB.108.014303} {\bibfield  {journal}
  {\bibinfo  {journal} {Phys. Rev. B}\ }\textbf {\bibinfo {volume} {108}},\
  \bibinfo {pages} {014303} (\bibinfo {year} {2023})}\BibitemShut {NoStop}%
\bibitem [{\citenamefont {Vanhala}\ and\ \citenamefont
  {Ojanen}(2023)}]{Vanhala2023}%
  \BibitemOpen
  \bibfield  {author} {\bibinfo {author} {\bibfnamefont {T.~I.}\ \bibnamefont
  {Vanhala}}\ and\ \bibinfo {author} {\bibfnamefont {T.}~\bibnamefont
  {Ojanen}},\ }\href {https://doi.org/10.1103/PhysRevResearch.5.033178}
  {\bibfield  {journal} {\bibinfo  {journal} {Phys. Rev. Res.}\ }\textbf
  {\bibinfo {volume} {5}},\ \bibinfo {pages} {033178} (\bibinfo {year}
  {2023})}\BibitemShut {NoStop}%
\bibitem [{\citenamefont {Jurcevic}\ \emph {et~al.}(2017)\citenamefont
  {Jurcevic}, \citenamefont {Shen}, \citenamefont {Hauke}, \citenamefont
  {Maier}, \citenamefont {Brydges}, \citenamefont {Hempel}, \citenamefont
  {Lanyon}, \citenamefont {Heyl}, \citenamefont {Blatt},\ and\ \citenamefont
  {Roos}}]{Jurcevic2017}%
  \BibitemOpen
  \bibfield  {author} {\bibinfo {author} {\bibfnamefont {P.}~\bibnamefont
  {Jurcevic}}, \bibinfo {author} {\bibfnamefont {H.}~\bibnamefont {Shen}},
  \bibinfo {author} {\bibfnamefont {P.}~\bibnamefont {Hauke}}, \bibinfo
  {author} {\bibfnamefont {C.}~\bibnamefont {Maier}}, \bibinfo {author}
  {\bibfnamefont {T.}~\bibnamefont {Brydges}}, \bibinfo {author} {\bibfnamefont
  {C.}~\bibnamefont {Hempel}}, \bibinfo {author} {\bibfnamefont {B.~P.}\
  \bibnamefont {Lanyon}}, \bibinfo {author} {\bibfnamefont {M.}~\bibnamefont
  {Heyl}}, \bibinfo {author} {\bibfnamefont {R.}~\bibnamefont {Blatt}}, \ and\
  \bibinfo {author} {\bibfnamefont {C.~F.}\ \bibnamefont {Roos}},\ }\href
  {https://doi.org/10.1103/PhysRevLett.119.080501} {\bibfield  {journal}
  {\bibinfo  {journal} {Phys. Rev. Lett.}\ }\textbf {\bibinfo {volume} {119}},\
  \bibinfo {pages} {080501} (\bibinfo {year} {2017})}\BibitemShut {NoStop}%
\bibitem [{\citenamefont {Flaschner}\ \emph {et~al.}(2018)\citenamefont
  {Flaschner}, \citenamefont {Vogel}, \citenamefont {Tarnowski}, \citenamefont
  {Rem}, \citenamefont {Luhmann}, \citenamefont {Heyl}, \citenamefont {Budich},
  \citenamefont {Mathey}, \citenamefont {Sengstock},\ and\ \citenamefont
  {Weitenberg}}]{Flaschner2018}%
  \BibitemOpen
  \bibfield  {author} {\bibinfo {author} {\bibfnamefont {N.}~\bibnamefont
  {Flaschner}}, \bibinfo {author} {\bibfnamefont {D.}~\bibnamefont {Vogel}},
  \bibinfo {author} {\bibfnamefont {M.}~\bibnamefont {Tarnowski}}, \bibinfo
  {author} {\bibfnamefont {B.~S.}\ \bibnamefont {Rem}}, \bibinfo {author}
  {\bibfnamefont {D.-S.}\ \bibnamefont {Luhmann}}, \bibinfo {author}
  {\bibfnamefont {M.}~\bibnamefont {Heyl}}, \bibinfo {author} {\bibfnamefont
  {J.~C.}\ \bibnamefont {Budich}}, \bibinfo {author} {\bibfnamefont
  {L.}~\bibnamefont {Mathey}}, \bibinfo {author} {\bibfnamefont
  {K.}~\bibnamefont {Sengstock}}, \ and\ \bibinfo {author} {\bibfnamefont
  {C.}~\bibnamefont {Weitenberg}},\ }\href
  {https://doi.org/10.1038/s41567-017-0013-8} {\bibfield  {journal} {\bibinfo
  {journal} {Nature Phys.}\ }\textbf {\bibinfo {volume} {14}},\ \bibinfo
  {pages} {265} (\bibinfo {year} {2018})}\BibitemShut {NoStop}%
\bibitem [{\citenamefont {de~Moura}\ and\ \citenamefont
  {Lyra}(1998)}]{Moura1998}%
  \BibitemOpen
  \bibfield  {author} {\bibinfo {author} {\bibfnamefont {F.~A. B.~F.}\
  \bibnamefont {de~Moura}}\ and\ \bibinfo {author} {\bibfnamefont {M.~L.}\
  \bibnamefont {Lyra}},\ }\href {https://doi.org/10.1103/PhysRevLett.81.3735}
  {\bibfield  {journal} {\bibinfo  {journal} {Phys. Rev. Lett.}\ }\textbf
  {\bibinfo {volume} {81}},\ \bibinfo {pages} {3735} (\bibinfo {year}
  {1998})}\BibitemShut {NoStop}%
\bibitem [{\citenamefont {Khan}\ \emph {et~al.}(2019)\citenamefont {Khan},
  \citenamefont {Lopes}, \citenamefont {Pires},\ and\ \citenamefont {dos
  Santos}}]{Niaz2019}%
  \BibitemOpen
  \bibfield  {author} {\bibinfo {author} {\bibfnamefont {N.~A.}\ \bibnamefont
  {Khan}}, \bibinfo {author} {\bibfnamefont {J.~M. V.~P.}\ \bibnamefont
  {Lopes}}, \bibinfo {author} {\bibfnamefont {J.~P.~S.}\ \bibnamefont {Pires}},
  \ and\ \bibinfo {author} {\bibfnamefont {J.~M. B.~L.}\ \bibnamefont {dos
  Santos}},\ }\href {https://doi.org/10.1088/1361-648x/ab03ad} {\bibfield
  {journal} {\bibinfo  {journal} {J. Phys.: Condens. Matter}\ }\textbf
  {\bibinfo {volume} {31}},\ \bibinfo {pages} {175501} (\bibinfo {year}
  {2019})}\BibitemShut {NoStop}%
\bibitem [{\citenamefont {Aubry}\ and\ \citenamefont
  {Andr{\'e}}(1980)}]{AAModel1980}%
  \BibitemOpen
  \bibfield  {author} {\bibinfo {author} {\bibfnamefont {S.}~\bibnamefont
  {Aubry}}\ and\ \bibinfo {author} {\bibfnamefont {G.}~\bibnamefont
  {Andr{\'e}}},\ }\href@noop {} {\bibfield  {journal} {\bibinfo  {journal}
  {Ann. Israel Phys. Soc.}\ }\textbf {\bibinfo {volume} {133}},\ \bibinfo
  {pages} {3} (\bibinfo {year} {1980})}\BibitemShut {NoStop}%
\bibitem [{\citenamefont {Anderson}(1958)}]{Anderson1958}%
  \BibitemOpen
  \bibfield  {author} {\bibinfo {author} {\bibfnamefont {P.~W.}\ \bibnamefont
  {Anderson}},\ }\href {https://doi.org/10.1103/PhysRev.109.1492} {\bibfield
  {journal} {\bibinfo  {journal} {Phys. Rev.}\ }\textbf {\bibinfo {volume}
  {109}},\ \bibinfo {pages} {1492} (\bibinfo {year} {1958})}\BibitemShut
  {NoStop}%
\bibitem [{\citenamefont {Balian}\ \emph {et~al.}(1984)\citenamefont {Balian},
  \citenamefont {Maynard},\ and\ \citenamefont {Toulouse}}]{Balian1984}%
  \BibitemOpen
  \bibfield  {author} {\bibinfo {author} {\bibfnamefont {R.}~\bibnamefont
  {Balian}}, \bibinfo {author} {\bibfnamefont {R.}~\bibnamefont {Maynard}}, \
  and\ \bibinfo {author} {\bibfnamefont {G.}~\bibnamefont {Toulouse}},\ }\href
  {https://doi.org/10.1142/0031} {\emph {\bibinfo {title} {Ill-Condensed
  Matter}}}\ (\bibinfo  {publisher} {NORTH-HOLLAND Publishing CO. (1984)},\
  \bibinfo {year} {1984})\BibitemShut {NoStop}%
\bibitem [{\citenamefont {Pires}\ \emph {et~al.}(2019)\citenamefont {Pires},
  \citenamefont {Khan}, \citenamefont {Lopes},\ and\ \citenamefont {dos
  Santos}}]{Pires2019}%
  \BibitemOpen
  \bibfield  {author} {\bibinfo {author} {\bibfnamefont {J.~P.~S.}\
  \bibnamefont {Pires}}, \bibinfo {author} {\bibfnamefont {N.~A.}\ \bibnamefont
  {Khan}}, \bibinfo {author} {\bibfnamefont {J.~M. V.~P.}\ \bibnamefont
  {Lopes}}, \ and\ \bibinfo {author} {\bibfnamefont {J.~M. B.~L.}\ \bibnamefont
  {dos Santos}},\ }\href {https://doi.org/10.1103/PhysRevB.99.205148}
  {\bibfield  {journal} {\bibinfo  {journal} {Phys. Rev. B}\ }\textbf {\bibinfo
  {volume} {99}},\ \bibinfo {pages} {205148} (\bibinfo {year}
  {2019})}\BibitemShut {NoStop}%
\bibitem [{\citenamefont {Khan}\ \emph {et~al.}(2020)\citenamefont {Khan},
  \citenamefont {Pires}, \citenamefont {Lopes},\ and\ \citenamefont {dos
  Santos}}]{Niaz2020}%
  \BibitemOpen
  \bibfield  {author} {\bibinfo {author} {\bibfnamefont {N.~A.}\ \bibnamefont
  {Khan}}, \bibinfo {author} {\bibfnamefont {J.~P.~S.}\ \bibnamefont {Pires}},
  \bibinfo {author} {\bibfnamefont {J.~M. V.~P.}\ \bibnamefont {Lopes}}, \ and\
  \bibinfo {author} {\bibfnamefont {J.~M. B.~L.}\ \bibnamefont {dos Santos}},\
  }\href {https://doi.org/10.1051/epjconf/202023305011} {\bibfield  {journal}
  {\bibinfo  {journal} {EPJ Web Conf.}\ }\textbf {\bibinfo {volume} {233}},\
  \bibinfo {pages} {05011} (\bibinfo {year} {2020})}\BibitemShut {NoStop}%
\bibitem [{\citenamefont {Khan}\ \emph {et~al.}(2022)\citenamefont {Khan},
  \citenamefont {Muhammad},\ and\ \citenamefont {Sajid}}]{Niaz2022}%
  \BibitemOpen
  \bibfield  {author} {\bibinfo {author} {\bibfnamefont {N.~A.}\ \bibnamefont
  {Khan}}, \bibinfo {author} {\bibfnamefont {S.}~\bibnamefont {Muhammad}}, \
  and\ \bibinfo {author} {\bibfnamefont {M.}~\bibnamefont {Sajid}},\ }\href
  {https://doi.org/10.1016/j.physe.2022.115150} {\bibfield  {journal} {\bibinfo
   {journal} {Physica E}\ }\textbf {\bibinfo {volume} {139}},\ \bibinfo {pages}
  {115150} (\bibinfo {year} {2022})}\BibitemShut {NoStop}%
\bibitem [{\citenamefont {Khan}(2023)}]{Niaz2023CJP}%
  \BibitemOpen
  \bibfield  {author} {\bibinfo {author} {\bibfnamefont {N.~A.}\ \bibnamefont
  {Khan}},\ }\href {https://doi.org/https://doi.org/10.1016/j.cjph.2023.07.011}
  {\bibfield  {journal} {\bibinfo  {journal} {Chin. J. Phys.}\ }\textbf
  {\bibinfo {volume} {85}},\ \bibinfo {pages} {733} (\bibinfo {year}
  {2023})}\BibitemShut {NoStop}%
\bibitem [{\citenamefont {Mason}\ and\ \citenamefont
  {Handscomb}(2002)}]{Mason2002}%
  \BibitemOpen
  \bibfield  {author} {\bibinfo {author} {\bibfnamefont {J.}~\bibnamefont
  {Mason}}\ and\ \bibinfo {author} {\bibfnamefont {D.}~\bibnamefont
  {Handscomb}},\ }\href {https://doi.org/0.1201/9781420036114} {\emph {\bibinfo
  {title} {Chebyshev Polynomials}}}\ (\bibinfo  {publisher} {CRC Press, New
  York (2002)},\ \bibinfo {year} {2002})\BibitemShut {NoStop}%
\bibitem [{Note1()}]{Note1}%
  \BibitemOpen
  \bibinfo {note} {The Hamiltonian and all energy scales can be normalized by
  dividing $2Dt+\protect \mathcal {F}$, where $D$ is the dimension of the
  system, $t$ is the hopping integral, and $\protect \mathcal {F}$ is a number
  that imposes the Hamiltonian spectrum to be in the interval $\left
  [-1,\protect \,1\right ]$.}\BibitemShut {Stop}%
\bibitem [{\citenamefont {Iacopi}\ \emph {et~al.}(2016)\citenamefont {Iacopi},
  \citenamefont {Boeckl},\ and\ \citenamefont {Jagadish}}]{Iacopi2016}%
  \BibitemOpen
  \bibfield  {author} {\bibinfo {author} {\bibfnamefont {F.}~\bibnamefont
  {Iacopi}}, \bibinfo {author} {\bibfnamefont {J.}~\bibnamefont {Boeckl}}, \
  and\ \bibinfo {author} {\bibfnamefont {C.}~\bibnamefont {Jagadish}},\ }\href
  {https://doi.org/978-0-12-804272-4} {\emph {\bibinfo {title} {2D
  Materials}}}\ (\bibinfo  {publisher} {Academic Press/Elsevier},\ \bibinfo
  {year} {2016})\BibitemShut {NoStop}%
\bibitem [{\citenamefont {Das~Sarma}\ \emph {et~al.}(1988)\citenamefont
  {Das~Sarma}, \citenamefont {He},\ and\ \citenamefont {Xie}}]{Sarma1988}%
  \BibitemOpen
  \bibfield  {author} {\bibinfo {author} {\bibfnamefont {S.}~\bibnamefont
  {Das~Sarma}}, \bibinfo {author} {\bibfnamefont {S.}~\bibnamefont {He}}, \
  and\ \bibinfo {author} {\bibfnamefont {X.~C.}\ \bibnamefont {Xie}},\ }\href
  {https://doi.org/10.1103/PhysRevLett.61.2144} {\bibfield  {journal} {\bibinfo
   {journal} {Phys. Rev. Lett.}\ }\textbf {\bibinfo {volume} {61}},\ \bibinfo
  {pages} {2144} (\bibinfo {year} {1988})}\BibitemShut {NoStop}%
\bibitem [{\citenamefont {Biddle}\ and\ \citenamefont
  {Das~Sarma}(2010)}]{Biddle2010}%
  \BibitemOpen
  \bibfield  {author} {\bibinfo {author} {\bibfnamefont {J.}~\bibnamefont
  {Biddle}}\ and\ \bibinfo {author} {\bibfnamefont {S.}~\bibnamefont
  {Das~Sarma}},\ }\href {https://doi.org/10.1103/PhysRevLett.104.070601}
  {\bibfield  {journal} {\bibinfo  {journal} {Phys. Rev. Lett.}\ }\textbf
  {\bibinfo {volume} {104}},\ \bibinfo {pages} {070601} (\bibinfo {year}
  {2010})}\BibitemShut {NoStop}%
\bibitem [{\citenamefont {Ganeshan}\ \emph {et~al.}(2015)\citenamefont
  {Ganeshan}, \citenamefont {Pixley},\ and\ \citenamefont
  {Das~Sarma}}]{Ganeshan2015}%
  \BibitemOpen
  \bibfield  {author} {\bibinfo {author} {\bibfnamefont {S.}~\bibnamefont
  {Ganeshan}}, \bibinfo {author} {\bibfnamefont {J.~H.}\ \bibnamefont
  {Pixley}}, \ and\ \bibinfo {author} {\bibfnamefont {S.}~\bibnamefont
  {Das~Sarma}},\ }\href {https://doi.org/10.1103/PhysRevLett.114.146601}
  {\bibfield  {journal} {\bibinfo  {journal} {Phys. Rev. Lett.}\ }\textbf
  {\bibinfo {volume} {114}},\ \bibinfo {pages} {146601} (\bibinfo {year}
  {2015})}\BibitemShut {NoStop}%
\end{thebibliography}%

\end{document}